\documentclass[
11pt, 
english, 
singlespacing, 
headsepline, 
]{MastersDoctoralThesis} 

\usepackage[utf8]{inputenc} 
\usepackage[T1]{fontenc} 

\usepackage{mathpazo} 

\usepackage{mathrsfs}
\usepackage{epstopdf}
\usepackage{amsmath}
\usepackage{amssymb}
\usepackage{subfigure}

\newcommand{\bo}{\raise-1mm\hbox{\Large$\Box$}}

\newcommand{\f}[2]{\frac{#1}{#2}}

\newcommand{\la}{\langle}
\newcommand{\ra}{\rangle}
\newcommand{\w}{\omega}
\newcommand{\kp}{\kappa}
\newcommand{\be}{\begin{equation}}
\newcommand{\ee}{\end{equation}}
\newcommand{\bea}{\begin{eqnarray}}
\newcommand{\eea}{\end{eqnarray}}

\usepackage{hyperref}
\usepackage{url}
\usepackage{xcolor}
\usepackage{color}
\definecolor{amaranth}{rgb}{0.9, 0.17, 0.31}
\definecolor{palatinateblue}{rgb}{0.15, 0.23, 0.89}
\definecolor{brightpink}{rgb}{1.0, 0.0, 0.5}
\definecolor{vividviolet}{rgb}{0.62, 0.0, 1.0}
\hypersetup{ linktoc=all,
    colorlinks, linkcolor={palatinateblue},
    citecolor={brightpink}, urlcolor={amaranth}}

\usepackage[style=phys,sorting=none]{biblatex}

\usepackage[autostyle=true]{csquotes} 

\thesistitle{A modification to the mirror trajectory corresponding to shell collapse} 
\supervisor{Michael\textsc{ R.R.Good}} 
\examiner{} 
\degree{Bachelor of Science in Physics} 
\author{Khalykbek Yelshibekov} 
\addresses{} 

\subject{Theoretical Physics} 
\keywords{} 
\university{\href{http://www.nu.edu.kz}{Nazarbayev University}} 
\department{\href{http://sst.nu.edu.kz}{School of Science and Technology}} 
\group{\href{https://sst.nu.edu.kz/en/physics}{Physics Department}} 
\faculty{\href{https://sst.nu.edu.kz/en/physics}{Physics Department}} 

\AtBeginDocument{
\hypersetup{pdftitle=\ttitle} 
\hypersetup{pdfauthor=\authorname} 
\hypersetup{pdfkeywords=\keywordnames} 
}

\begin{document}

\frontmatter 

\pagestyle{plain} 


\begin{titlepage}
\begin{center}

\vspace*{.06\textheight}
{\scshape\LARGE \univname\par}\vspace{1.5cm} 
\textsc{\Large Bachelor Thesis}\\[0.5cm] 

\HRule \\[0.4cm] 
{\huge \bfseries \ttitle\par}\vspace{0.4cm} 
\HRule \\[1.5cm] 
 
\begin{minipage}[t]{0.4\textwidth}
\begin{flushleft} \large
\emph{Author:}\\
\href{mailto:khalykbek.yelshibekov@nu.edu.kz}{\authorname} 
\end{flushleft}
\end{minipage}
\begin{minipage}[t]{0.4\textwidth}
\begin{flushright} \large
\emph{Supervisor:} \\
\href{https://michaelrrgood.weebly.com/}{\supname} 
\end{flushright}
\end{minipage}\\[3cm]
 
\vfill

\large \textit{A thesis submitted in fulfillment of the requirements\\ for the degree of \degreename}\\[0.3cm] 
\textit{in the}\\[0.4cm]
\groupname\\\deptname\\[2cm] 
 
\vfill

{\large \today}\\[4cm] 

\vfill
\end{center}
\end{titlepage}


\begin{declaration}
\addchaptertocentry{\authorshipname} 
\noindent I, \authorname, declare that this thesis titled, \enquote{\ttitle} and the work presented in it are my own. I confirm that:

\begin{itemize} 
\item This thesis is drawn almost exclusively from \cite{horizonless} and \cite{GTC}.
\item This work was done wholly or mainly while in candidature for a research degree at this University.
\item Where any part of this thesis has previously been submitted for a degree or any other qualification at this University or any other institution, this has been clearly stated.
\item Where I have consulted the published work of others, this is always clearly attributed.
\item I have acknowledged all main sources of help.
\end{itemize}
 
\noindent Signed:\\
\rule[0.5em]{25em}{0.5pt} 
 
\noindent Date:\\
\rule[0.5em]{25em}{0.5pt} 
\end{declaration}


\vspace*{0.2\textheight}

\noindent\enquote{\itshape I'm smart enough to know that I'm dumb.}\bigbreak

\hfill Richard Feynman


\begin{abstract}
\addchaptertocentry{\abstractname} 
The well-known moving mirror trajectory corresponding to black hole shell collapse scenario (BHC) appears to produce infinite amount of energy. By changing the final velocity of the BHC mirror from speed of light to a free variable $\xi$ a new mirror trajectory was found. This modified BHC mirror produce finite energy and emits thermal radiation. In the limit of $\xi \to c$ the BHC mirror is restored. In this thesis the energy and particle production of the modified BHC mirror as well as the correlations in its radiation are presented.
\end{abstract}




\tableofcontents 


\dedicatory{Dedicated to Mama Roza} 


\mainmatter 

\pagestyle{thesis} 

\chapter{Introduction}
\label{sec:intro}

\section{What is a moving mirror?}

A moving mirror is a boundary:
\be \Psi|_{z(t)} =0 \label{eq:mirror_def} \ee
Here $z(t)$ is a trajectory of the mirror in (1+1) dimensional spacetime, and $\Psi$ is a massless scalar field. As you can see, this equation works for all type of mirrors. 

For example, perfect electric conductors (PEC) have electric field $E=0$. That's why people use silver (one of the best conducting materials) as mirrors for light. Another good example is an infinite square well.

\begin{figure}[ht]
\centering
\includegraphics[width=2.5in]{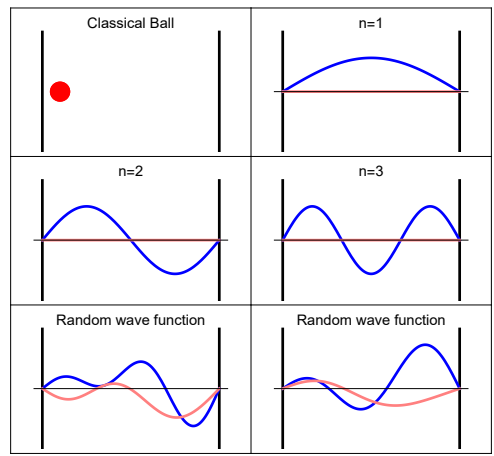}
\caption{Particle in a box}
\end{figure}
Here again we have wave function $\Psi=0$ at the boundaries. So you can think of the moving mirror as a moving infinite potential well. 

Please be aware of some popular misconceptions about moving mirrors. Note that the delta function potential is not a mirror. Despite being infinite at the center, the delta function potential can have a non-zero wave function \textit{inside} of it. Therefore particles can tunnel through the delta function potential, but they cannot tunnel through a mirror defined by Eq.~(\ref{eq:mirror_def}). That is why they are sometimes called perfectly reflecting mirrors. 

In the scope of this thesis we are interested in moving mirrors defined by Eq.~(\ref{eq:mirror_def}), where $\Psi$ is a massless scalar field. From QFT we know that such kind of fields can be described by Klein–Gordon equation (which is an analog of Schrödinger equation in QFT):
$$ \Box \Psi=0 $$
The \textit{box} is also known as d'Alembert operator $\Box=-\frac{\partial^2}{c^2\partial^2t}+\frac{\partial^2}{\partial^2x}+\frac{\partial^2}{\partial^2y}+\frac{\partial^2}{\partial^2z}$.

\section{Why do we study moving mirrors?}

We want to study Hawking's radiation \cite{Hawking2}, which is a quantum effect in the presence of strong gravity. There is a huge problem with black holes, and that is an \textit{information paradox} \cite{Hawking3}. Where did the information that falls into a black hole disappear to \cite{Wolf,0808.2096,1111.6580,1301.4504,0808.2096,1111.6580,1301.4504}?  Attempts to recover information from evaporating black holes continue to produce new paradoxes, such as the firewall controversy \cite{amps, apologia, sam, naked}. 

In fact, a central piece of the puzzle regarding information loss, is the understanding of entanglement entropy, $S$. Let us consider the formation of a black hole from a gravitational collapse of some matter in a pure state. It is often believed that the entanglement entropy of the Hawking radiation received at null infinity, which should be zero at the beginning before any radiation arrives, should first increase, but then decrease at some point (known as the ``Page time''), so that eventually the entanglement entropy vanishes. Such a ``Page curve'' \cite{page1, page1b, page2} --- the plot of the entanglement entropy against time --- is crucial, since it gives insight into how information may be retrieved from the highly scrambled Hawking emission. 

There are, however, subtleties that are often overlooked. Notably, any calculation of entanglement entropy necessitates regularizing ultraviolet divergences (the kind that do not affect the spin-statistics connection \cite{Good:2012cp}). One notices that imposing a cutoff is a tricky procedure since modes that have sufficiently high energy at some point in the spacetime can be red-shifted at some other point due to spacetime curvature. This implies that a mode that is beyond a cutoff scale can be red-shifted below the scale, so the cutoff is not well-defined \cite{0901.3156, 9403137}. 

Furthermore, the results obtained could depend on the cutoff scheme. Progress has been made recently with the introduction of the causal-splitting regularization scheme of Bianchi and Smerlak \cite{Bianchi:2014vea, Bianchi:2014qua,1409.0144}.  The method allows one to compute the production of entanglement entropy in a \emph{cutoff-independent} manner. However, even then, there are still a few puzzles regarding the entanglement entropy of an evaporating black hole.

Studying moving mirrors will likely help us with developing a theory of quantum gravity \cite{huh}. A central advantage to the moving mirror model is its simplicity.  This is both in general, and in the context of the recent one-to-one correspondence with a black hole \cite{Good:2016oey, paper1, paper2, Good:2016bsq}, which found that for one concrete example, the particle production is exactly the same in both the mirror and black hole cases in (1+1)-dimensions.  It is good to emphasize that one should not hope that a moving mirror model can fully resolve the information paradox of a \emph{bona fide} black hole, (like a rotating black hole with temperature $2\pi T = g - k$ \cite{Good:2014uja}, much less an extremal black hole e.g. \cite{Good:2008yd}), but understanding the subtleties of quantum field theory in a moving mirror model \emph{is} a first step toward the more complicated physics of black holes.  The simplicity of (1+1)-dimensions allows the crux of Hawking radiation to stand out more clearly, separated from the specialized details associated with higher dimensional curved geometry and back-scattering. With the one-to-one correspondence, the moving mirror model can be treated as an even more precise analogy to Hawking's original argument, and therefore it is of interest to extend the one-to-one correspondence to more physically realistic circumstances, while holding on to this fortunate simplicity.  However, even as a relatively simple theoretical model of black hole evaporation in (1+1) -dimensions, the moving mirror model, has in practice, been very hard to extend to solutions for exact trajectories where the global Bogoliubov coefficients may be evaluated. Few solutions have been found\footnote{For example: the  case of uniform acceleration of Davies-Fulling \cite{Davies:1977yv}; and the case of eternal thermal emission of Carlitz-Willey \cite{Carlitz:1986nh}.} and finite-nonzero-energy cases are scarce.\footnote{See the first known solution found by Walker-Davies \cite{Walker:1982} and the asymptotically static case in Good-Anderson-Evans \cite{Good:2013lca} and a drifting case in Good-Ong \cite{Good:2015nja}.} Nevertheless, mirror trajectories that produce finite amount of energy are precisely those that are physically more realistic. We are therefore interested in such trajectories.  It is important to recognize that the main model in this work is an extension \cite{Good:2018,Myrzakul2018} of the particular moving mirror which has a one-to-one correspondence to the exactly solvable black hole case \cite{Good:2016oey}. There are other recent extensions \cite{Good:2018aer,Good:2017ddq,Good:2017kjr,Good:2016yht}, including interesting uniformly accelerated trajectories which may be explored in 3+1 dimensions \cite{Proceedings}.  

\begin{figure}[ht]
\centering
\includegraphics[width=2.5in]{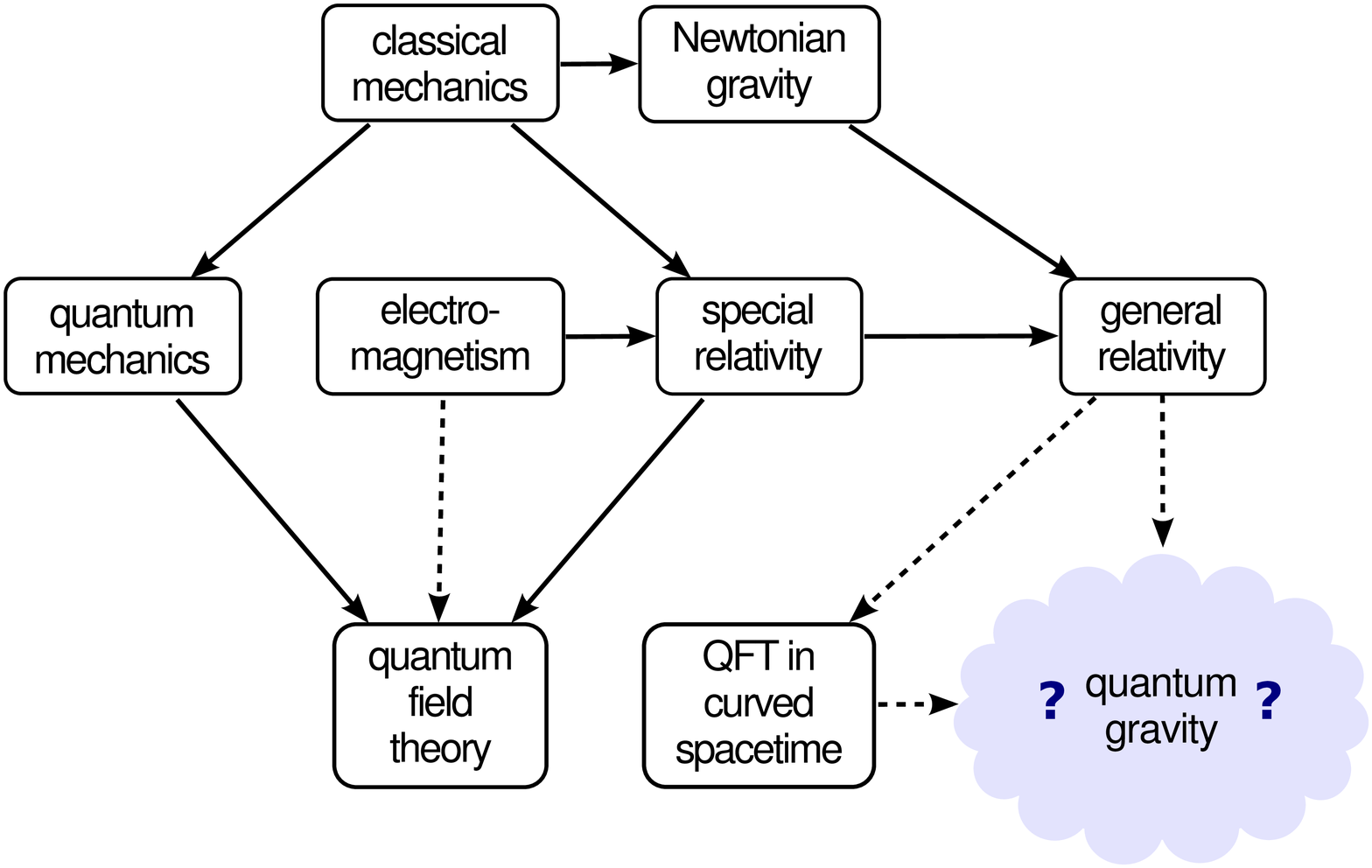}
\caption{Diagram showing the place of quantum gravity in the hierarchy of physics theories}
\end{figure}

By Einstein's equivalence principle, gravity and acceleration are \textit{indistinguishable}. That is, a free falling and freely floating in space with a constant velocity are indistinguishable for an observer experiencing them. Same with standing on the surface of the Earth and accelerating with $a=9.8$ m/s$^2$ in the opposite direction.

Therefore, it is plausible, at least in a certain context, to do a kind of mapping from the Schwarzschild metric to flat spacetime by simply moving with a corresponding acceleration. In this case the black hole center will be moving with respect to you, appearing as a moving mirror, and emitting energy (as Hawking radiation).

This is another popular misconception about moving mirrors. The mirror actually corresponds to the center of the black hole, and not its event horizon.

Throughout this paper we work in (1+1) dimensional spacetime. In this framework the whole space is divided into left and right sides. Please keep in mind that when we talk about about left and right sides of the mirror we usually mean a very distant observer to the left and to the right correspondingly. Moreover, we use specific set of units $G = c=\hbar=k_B=1$.

\chapter{The Machinery of Moving Mirrors} 

\label{sec:background}

In Section (\ref{sec:somebackground}), we shall first introduce some basic concepts necessary for understanding our construction of a horizonless solution that generalizes the ``black mirror''.   We then discuss the removal of the acceleration horizon in Section (\ref{sec:removehorizon}).  The solution satisfies three criteria: the presence of a horizonless temperature, an appropriately scaled acceleration parameter, and the termination of evaporation at late times.  

\section{Some Conventions of Moving Mirrors} \label{sec:somebackground} 
As the simplest example of the dynamical Casimir effect, the moving mirror model also serves as a way to understand black hole evaporation by imposing an external boundary condition in 1+1 dimensions on the quantum field, rather than an external curved spacetime. Consider then, such a boundary that does not accelerate forever, starting and ending at time-like past infinity $i^-$, and time-like future infinity $i^+$, respectively; possessing  asymptotically zero acceleration in both the far past and far future, and always moving slower than the speed of light.  This fully asymptotically inertial mirror will contain no horizon.  Thus, it will contain no pathological acceleration singularity either.  The plot of such a trajectory is in Figure (\ref{fig:Penrose}). A salient pay-off for horizon-removal is that the mirror system, in addition to being unitary (henceforth, by unitarity we always mean unitarity in the broad sense allowed by Bianchi-Smerlak criteria), produces only a finite amount of total energy, as we will demonstrate in Section (\ref{sec:energy}).

  
The quantum field, $\Psi$, is the massless scalar of the Klein-Gordon equation, (for a nonlinear investigation of the KG equation in the more general context of quantum field theory under external conditions, see \cite{Xiong:2014oga, Good:2014}), $\Box \Psi = 0$, whose value is zero, $\Psi|_z =0$, when evaluated at the position of the moving mirror, $z(t)$. The modes, $\phi_{\w'}$ and $\psi_\w$, are on equal footing in the sense that they can both be used to expand the field:
\be \Psi = \int_0^\infty d\w'\; \left[a_{\w'} \phi_{\w'} + a^\dagger_{\w'}\phi^*_{\w'}\right] = \int_0^\infty d\w\; \left[b_\w \psi_\w + b^{\dagger}_\w\psi^{*}_\w\right]. \ee
The modes are orthonormal and complete and can be exactly solved in the (1+1)-dimensional case:
\be 
\phi_{\w'} = (4\pi \w')^{-1/2} [e^{-i\w' v } - e^{-i\w' p(u)} ],
\ee
\be
\psi_{\w}  = (4\pi \w)^{-1/2} [e^{-i\w f(v)} - e^{-i\w u} ],
\ee
where the functions $p(u)$ and $f(v)$ are the usual notation for the ray-tracing functions, which are intimately related to the trajectory of the mirror $z(t)$ itself, see \cite{Good:2013lca} and Section (\ref{sec:fourusefulfunctions}). The famous Bogoliubov coefficients appear by expanding one set of modes in terms of the other set of modes, 
\be \label{phichi} \phi_{\w'} = \int_0^\infty d\w\; \left[\alpha_{\w'\w} \psi_{\w} + \beta_{\w'\w}\psi^*_{\w}\right], \ee
\be \label{chiphi} \psi_{\w} = \int_0^\infty d\w'\; \left[\alpha^*_{\w'\w} \phi_{\w'} - \beta_{\w'\w}\phi^*_{\w'}\right], \ee
where  
\be \alpha_{\w'\w} = (\phi_{\w'},\psi_{\w}), \qquad \beta_{\w'\w} = -(\phi_{\w'},\psi^*_{\w}), \ee
with the flat space scalar product defined in null coordinates, $(u,v)$, by 
\be (\phi_{\w'} , \psi_{\w} ) \equiv i \int_{-\infty}^{\infty} d u\; \phi_{\w'}^* \stackrel{\leftrightarrow}{\partial_u} \psi_{\w} + i \int_{-\infty}^{\infty} d v\; \phi_{\w'}^* \stackrel{\leftrightarrow}{\partial_v} \psi_{\w}. \ee
The Bogoliubov coefficients $\alpha_{\w'\w}$ and $\beta_{\w'\w}$ also give the operators $a_{\w'}$ and $a_{\w'}^\dagger$ in terms of the operators $b_{\w}$ and $b^{\dagger}_{\w}$, while the orthonormality of the modes hold according to the usual convention, see \cite{Good:2015nja} for more detail. 
%

\section{Four Trajectory Functions of Mirror Physics}\label{sec:fourusefulfunctions}
There are four functions, $v_s(u)$, $u_s(v)$, $x_s(t)$, $t_s(x)$, which are useful for doing global calculations involving the aforementioned field modes.  The first two are the ray-tracing functions (expressed in null coordinates), where $u_s(u) \equiv p(u)$ and $v_s(v)\equiv f(v)$, and the last two are the associated spacetime coordinate functions.  The inverses are expressed like so: 
\be u_s(v) = v_s^{-1}(u),\quad x_s(t) = t_s^{-1}(x).\ee
We shall collectively call all four of them, ``shock wave functions'' for short, after the collapse of the null shell shock wavefront description to form a black hole.  They are just the trajectory functions of the mirror in different coordinates, null or spacetime.  There are many other auxiliary functions, such as $t_s(v)$, $v_s(t)$, $t_s(u)$, $u_s(t)$. However, the original four functions of coordinates $v$, $u$, $t$, and $x$ will prove efficient at calculating observables.  The information about how the field modes become red-shifted due to external conditions is fully contained in these four functions. The relationships between them are demonstrated as follows. 

First consider the usual null coordinates on Minkowski spacetime $ u \equiv t-x$ and $ v \equiv t+x$, and their analogous auxiliary functions as functions of time, 
\be u_s(t) = t- x_s(t), \quad v_s(t) = t+ x_s(t). \label{coordshock} \ee
These contain the shock function $x_s(t)$, which is the trajectory of the mirror.  The inverses of Eqs.~(\ref{coordshock}), contain the shock functions, $u_s(v)$ and $v_s(u)$,
\be t_s(u) = \frac{1}{2}(v_s(u) + u), \quad t_s(v) = \frac{1}{2}(u_s(v) + v). \ee
Functional inverses should be obvious from the notation.  Useful auxiliary inverses are: $t_s(u) = u_s^{-1}(t)$, and $t_s(v) = v_s^{-1}(t)$. The total energy emitted, the energy flux, and the beta Bogoliubov coefficients have expressions that are conveniently written in terms of the four shock functions.
    
\section{Connection to Collapse Geometry}
Now I would like to indicate, following Unruh \cite{Unruh:1976db} a precise connection between the mirror model and collapse geometry, and summarize the resulting correlation functions and radiation flux.

Consider for simplicity a spherically symmetric collapsing shell of matter. We have vacuum inside and outside the shell, while the shell carries a given amount of mass (and possibly other quantum numbers). Thanks to Birkhoff’s theorem, we know the metric in both regions:

\be ds^2 = 
    \left\{
        \begin{array}{ll}
            -d\tau^2 + dr^2 + r^2 d\Omega^2 & \textnormal{, for } \tau+r < V_s \\
            -\lambda^2 dt^2 + \lambda^{-2} dr^2 + r^2 d\Omega^2  & \textnormal{, for } t+r > v_s \\
        \end{array} 
    \right. \ee
Note that in order to exhibit the metric in each region in its familiar (static) form, two different sets of coordinates had to be used. It is convenient to introduce light-cone coordinates in each region. In the interior region we use simply $U=\tau-r$ and $V=\tau+r$, whereas in the outer region we first define the tortoise-coordinate $r_*$ through
\be \label{eq:dr} \frac{dr_*}{dr}=\frac{1}{\lambda^2} \ee
and then take $u=t-r_*$ and $v=t+r_*$ as light-cone coordinates. The space-time is described by the metric:

\be \label{eq:metric} ds^2 = 
    \left\{
        \begin{array}{ll}
            -dUdV + r^2 d\Omega^2 & \textnormal{, for } V < V_s \\
            -\lambda^2 dudv + r^2 d\Omega^2  & \textnormal{, for } v > v_s \\
        \end{array} 
    \right. \ee
where $r$ is determined through the relations
\be \label{eq:UVrelations} \begin{array}{ll}
            V-U=2r & \textnormal{, for } V < V_s \\
            v-u=2r_*(r)  & \textnormal{, for } v > v_s \\
        \end{array} \ee
        
When we paste together the two coordinate systems for the interior and exterior region to form a global coordinate-system, we can choose to coincide with Eq.~(\ref{eq:metric}) either in the exterior or in the interior region.  The first choice is natural from the point of view of a distant observer, while the second is more convenient to implement the boundary condition at the origin and to display the complete space-time structure.

Let us consider first the former choice, that is using $u$-$v$-coordinates in both regions and looking for a satisfactory coordinate-transformation $U(u)$ and $V(v)$. In the infinite past the space-time is flat and there is no difference between the two coordinate systems.  This implies that we can choose $V(v) =v$.  We find the function $U(u)$ by demanding that along the worldline $v=v_s$ of the shell the coordinate $r$ should agree in both systems, because it has a gauge-invariant meaning (it determines the area of a two-sphere at constant radius and time). Applying Eq.~(\ref{eq:UVrelations}) along $v=v_s$ we obtain the implicit relation:
\be \label{eq:r*} r_*\left( r=\frac{v_s-U(u)}{2}\right)=\frac{v_s-u}{2} \ee
Differentiating this equation along the worldline of the shell we find, with the help of the defining Eq.~(\ref{eq:dr}) for $r_*$,
\be \frac{dU}{du}=\lambda^2(u,v_s) \ee
so that the metric becomes:
\be ds^2 = 
    \left\{
        \begin{array}{ll}
            -\lambda^2(u,v_s)dudv + r^2 d\Omega^2 & \textnormal{, for } v < v_s \\
            -\lambda^2(u,v) dudv + r^2 d\Omega^2  & \textnormal{, for } v > v_s \\
        \end{array} 
    \right. \ee
which is continuous along $v_s$. The metric is, of course, only valid for non-negative values of $r$, i.e. for $v\ge U(u)$. The world-line of the origin is therefore described by
\be v_0(u)=U(u) \ee
Since nothing can go beyond the regular origin, i.e.to negative $r$, it acts like a perfectly reflecting mirror.

In the $u$-$v$-frame the shell never crosses the horizon since $r_*$ and $t=v_s-r_*$ diverge as the horizon is approached. On the other hand we know that the shell reaches the origin in finite proper time. In order to describe the whole space-time, including the interior of the black hole it is convenient to use the $U$-$V$-coordinates, which provide a complete cover since they contain the origin until the shell reaches it. The space-time is then described by
\be ds^2 = 
    \left\{
        \begin{array}{ll}
            -dUdV + r^2 d\Omega^2 & \textnormal{, for } v < v_s \\
            -\lambda^2(u,v) \lambda^{-2}(u,v_s) dUdV + r^2 d\Omega^2  & \textnormal{, for } v > v_s \\
        \end{array} 
    \right. \ee
In spite of its appearance, the metric is regular on the horizon where $\lambda^2= 0$. The origin is stationary at $V=U$ until the shell reaches it.

For a shell of mass $M$ one has explicitly for the tortoise coordinate
\be r^* = r + 2 M \ln \left(\frac{r}{2M}-1\right)\ee
and thus from Eq.~(\ref{eq:r*}),
\be u = U - 4M \ln [(v_s-4M - U)/2] \ee

\section{How to Remove a Horizon}\label{sec:removehorizon}

There is an easy way to remove the horizon, (recall that $c=1$),
\be \lim_{t\rightarrow +\infty}| \dot{\mathfrak{z}}(t)| = 1,\ee
 from a future asymptotically null moving mirror trajectory, $\mathfrak{z}(t)$.  The idea is to modify this so that the new trajectory, $z(t)$, has
\be \lim_{t\rightarrow +\infty}| \dot{z}(t)| = \xi, \ee
where $0<\xi < 1$ is the future asymptotically drifting speed.  This can be achieved by writing the horizonless trajectory, $z(t)$, in terms of the trajectory with a horizon (henceforth `horizon trajectory'), $\mathfrak{z}(t)$, 
\be z(t) =  \xi  \mathfrak{z}(t). \ee
This works if 
\be \dot{z}(t) = \xi \dot{\mathfrak{z}}(t). \ee
Taking this approach helps answer whether the particle spectra can (1) reach equilibrium for an extended period of time, and (2) proceed to shut off.  The mirror does not strictly have a null horizon, yet as we will see, it can still achieve a ``thermal plateau'' (i.e. the emission is virtually thermal for some arbitrary finite amount of time).  This approach also ensures (3) the correct scale for the acceleration parameter $\kappa$ (not to be confused with the physical acceleration, see below).  A correct scale is critical for the red-shifting of the modes to correspond to the exactly solvable black hole case \cite{Good:2016oey, paper1, paper2, Good:2016bsq} in the limit $\xi \rightarrow 1$.  This automatically extends the mirror in the black hole-moving mirror correspondence by promoting it to a more physical footing where the total evaporation energy is finite and unitarity is preserved.  

While we have found the mirror solution that meets these strict requirements, a possible black hole counterpart calculation is beyond the scope of this work.  In the model we are about to present, we do not claim that it actually corresponds to any realistic evaporating black hole spacetime.  For the present work we only seek a simple mirror model in which the three conditions presented above are met, so that we may study the energy and energy flux, the entropy, the correlations, and the particle spectra, together in the absence of a horizon.  It may or may not have an exactly tractable black hole correspondence. In a subsequent work this will be investigated, but as we have emphasized in the Introduction, even if it has such a black hole correspondence, the absence of horizon in the mirror model does not necessary entail the absence of any event horizon or trapped surface in the black hole geometry.
\emph{It is worth pointing out that this mirror solution is new} --- it is the first explicit demonstration of a unitary solution with a thermal plateau\footnote{Other unitary plateau investigations exist, see Section (\ref{sec:conclusions}) for a discussion of one.} that has limiting red-shifting functions which correspond to the black hole-moving mirror system in \cite{Good:2016oey}.

The information contained in the trajectory equation of motion of the mirror is also contained in the shock functions.  The exactly solvable mirror case in \cite{Good:2016oey} has shock functions:
\bea
 v_{s}(u) &=& v_H - \kp^{-1} W\left[e^{\kp(v_H-u)}\right],\\
 u_{s}(v) &=& v - \kp^{-1} \ln \left[\kp(v_H - v)\right], \label{matchingfunction} \\
 x_{s}(t) &=& v_H -t - (2\kp)^{-1} W \left[ 2 e^{2\kp(v_H-t)} \right], \\
 t_{s}(x) &=& v_H -x -\kp^{-1} e^{2\kp x}.
 \eea
The $W$ is the product log or $W$ Lambert function, which commonly appears in thermal equilibrium contexts, e.g. Wien's law.\footnote{The maximum frequency of the (3+1)-dimensional Planck distribution, $\frac{ V \hbar }{\pi^2 c^3} \frac{\omega^3}{e^{\beta \hbar \omega} -1}$, is $\beta \hbar \omega_{\textrm{max}} = 3 + W\left[-\frac{3}{e^3}\right]$, i.e. the famous displacement law $\beta \hbar \omega_{\textrm{max}} = 2.82144$. }  One way to get these is as follows: Firstly, one has the simple form $u_s(v)$ as it is a simple choice for redshifting ray-tracing $f(v)$ function in the mirror case (or from the spacetime matching solution in the null-shell case). This is given.  Secondly, one takes the inverse to get $v_s(u)$.  While easy, as it turns out, it was unhelpful in obtaining the other shock wave functions. The efficient approach is to notice that $u_s(v)$ has a simpler form than $v_s(u)$, so one uses $u_s(v)$ again to write down $t_s(v)$.  The inverse of this can be calculated.  It is, of course, $v_s(t)$. (Note that if one chooses $v_s(u)$ to write down $t_s(u)$ instead, the inverse is not quite as straight-forward to compute, in fact it is much more complicated.)  So, using $v_s(t)$, one is set to write down $x_s(t) = v_s(t) - t$.  Its inverse is, fortuitously tractable, and gives the above expression for $t_s(x)$.  

We shall interchangeably call the horizon trajectory in the mirror analog case the ``black mirror'' \cite{Good:2016bsq} or ``BHC'' for short \cite{Good:2016oey}. The acceleration parameter $\kappa$ in the black mirror case can be identified with the surface gravity in the black hole case, $\kappa = (4M)^{-1}$, for all times. 


The new moving mirror has the following more complicated shock functions:

\be v_s(u) = \frac{2 \xi}{1+\xi}v_H + \frac{1-\xi}{1+\xi}u- \frac{\xi}{\kp}  W\left[\frac{2 e^{\frac{2\kp(v_H-u)}{1+\xi}}}{1+\xi}\right] ,\ee
\be u_s(v) = -\frac{2 \xi }{1-\xi }v_H + \frac{ 1 + \xi }{1-\xi }  v + \frac{\xi}{\kp} W\left[\frac{2 e^{\frac{2\kp( v_H - v)}{1-\xi}}}{1-\xi}\right], \label{matchingfunction2}\ee
\be x_s(t) = \xi  \left(v_H -t -\frac{W\left[ 2 e^{2\kp(v_H-t)} \right]}{2\kp}\right) \label{traj}, \ee
\be t_s(x) = v_H-\frac{x}{\xi } -\frac{1}{\kp} e^{2\kp x/\xi} .\ee


While these expressions still depend on the primary parameter $\kp$, the intricacy of these expressions arises from the introduction of a second parameter, $\xi$.  Recall that $v_H$ in the black mirror case is the location of the horizon.  We retain $v_H$ for completeness, but make no mistake: the mirror no longer asymptotes to infinite acceleration at a null horizon, located at $v_H$.  
We shall therefore refer to $v_H$ as a ``residual horizon''.  This mirror begins at rest in the far past, and therefore has no initial asymptotic horizon either.  The absence of horizons generates the finite total energy, akin to the notion that evaporating black holes exhale only a finite energy flux, consistent with the idea of conservation of energy. 

\chapter{The MBHC Trajectory} 

\label{sec:trajectory}

The motion of the mirror is given by the trajectory Eq.~(\ref{traj}),

\be \label{trajectory} z(t) =  -\xi  \left(\frac{1}{2\kappa} W\left[ 2 e^{-2\kappa t} \right]+t\right), \ee
where $v_H=0$ for simplicity, and $0<\xi<1$ is the final speed of the mirror as $t\rightarrow\infty$.  The motion is initially asymptotically static, $\lim_{t\rightarrow -\infty} \dot{z}(t) = 0$,
and most notably, the mirror does not approach a future asymptotically static resting state because its future asymptotic speed is
\be \lim_{t\rightarrow +\infty}| \dot{z}(t)| = \xi, \ee
making this trajectory future \emph{asymptotically coasting}.  The future drifting feature of this mirror means it is an exact model for a remnant \cite{Good:2015nja,Chen:2014jwq} as described by an early anticipation of such solutions by Wilczek in \cite{Wilczek:1993jn}. 

\begin{figure}[ht]
\centering
\mbox{\subfigure{\includegraphics[width=2.5in]{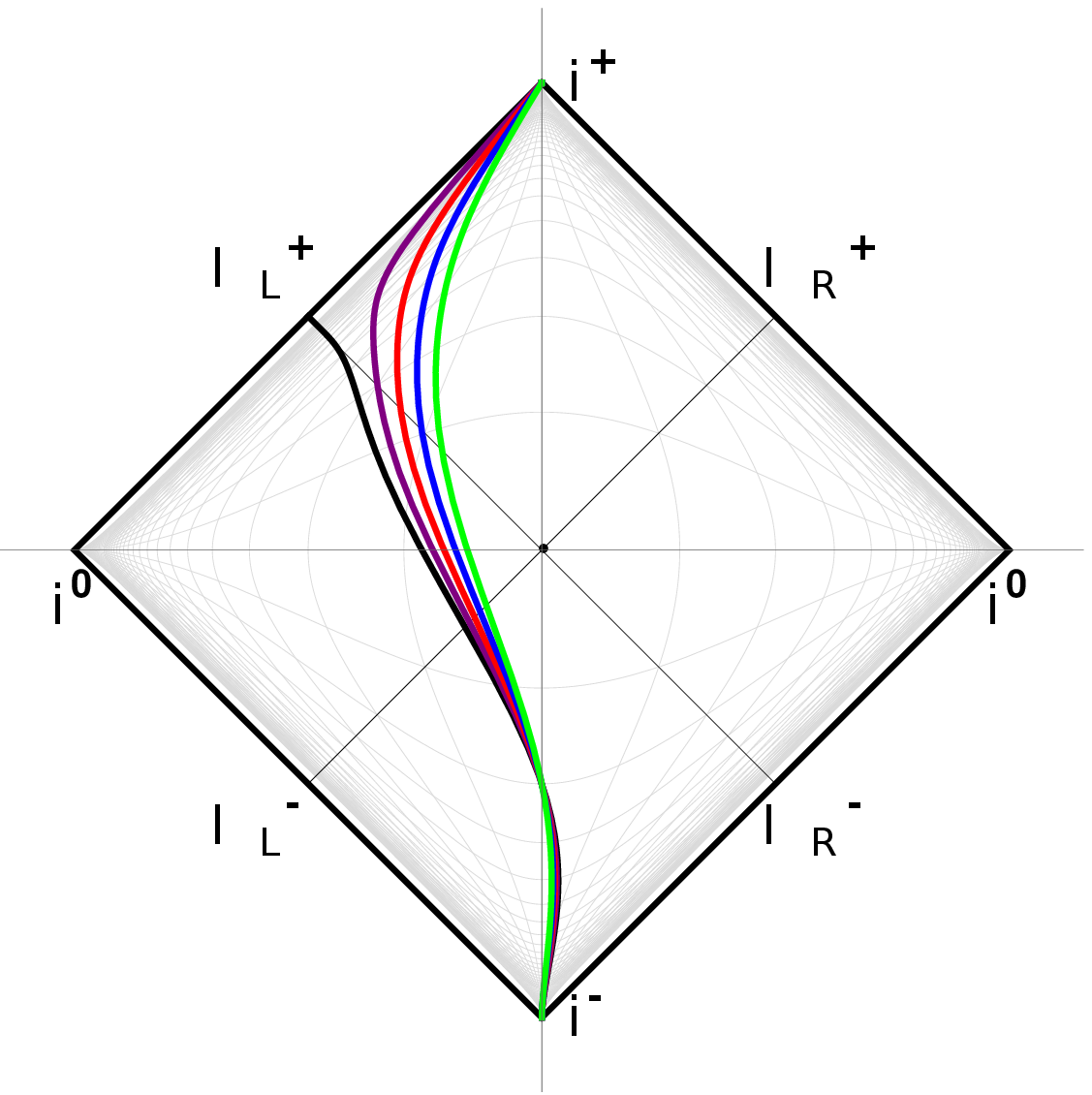}}\quad
\subfigure{\rotatebox{90}{\includegraphics[width=2.5in]{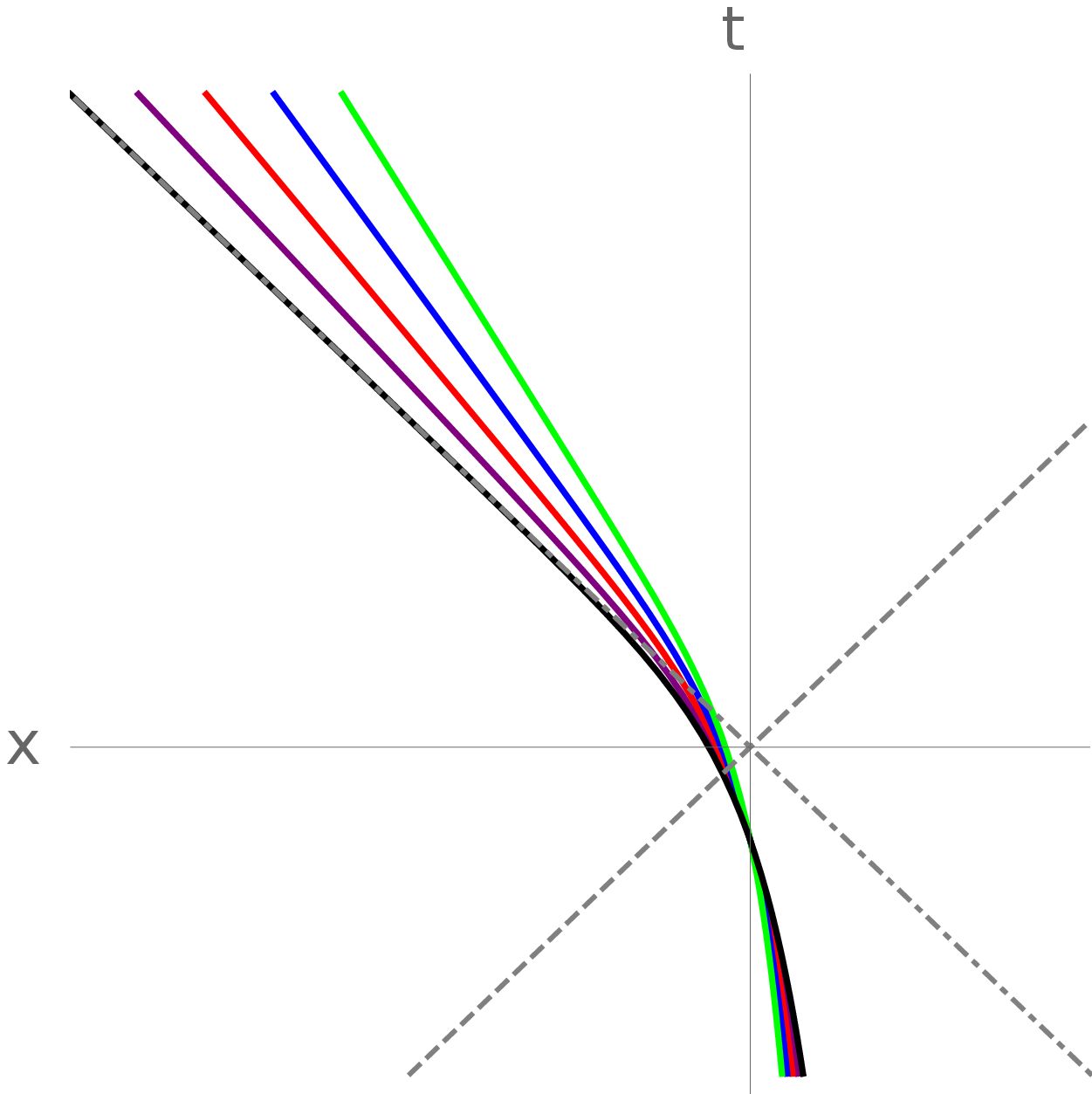} }}}
\caption{\textbf{Left:} In this Penrose diagram, the color curves are ``MBHC'', with asymptotically inertial trajectories.  The black curve is ``BHC'',
 a horizon mirror (moving mirror with a horizon) \cite{Good:2016oey}.  The different coasting speeds correspond to $\xi =  0.6$, $0.7$, $0.8$, $0.9$, for green, blue, red, purple, respectively. \textbf{Right:} The asymptotically inertial trajectories (MBHC) with the same final coasting speeds displayed in the usual spacetime diagram.  The dashed lines represent the light cone, and the dotted-dashed horizon line is at $v_H=0$.  The trajectory example here is the same as in the conformal diagram.  For comparison, the black line indicates the horizon mirror (BHC) \cite{Good:2016oey}, which contains a horizon coinciding with the light cone. \label{fig:Penrose}} 
\end{figure}

The trajectory Eq.(\ref{trajectory}) is plotted in both the spacetime and Penrose diagrams in Figure (\ref{fig:Penrose}).  The acceleration parameter, $\kp$, is $\kp>0$, and to be clear, it is \emph{not} the acceleration of the mirror, $\alpha(t) \neq \kappa$.  The rectilinear proper acceleration, $\alpha = \gamma^3\ddot{z}$, is time-dependent:
\be \label{propacc} \alpha(t) = -\frac{2 \kappa  \xi  W\left(2 e^{-2 \kappa  t}\right)}{\left(W\left(2 e^{-2 \kappa  t}\right)+1\right)^3 \left(1-\frac{\xi ^2}{\left(W\left(2 e^{-2 \kappa  t}\right)+1\right)^2}\right)^{3/2}}. \ee
The negative sign on Eq.(\ref{propacc}) gives a mirror whose motion is to the left.  The acceleration has asymptotic behavior such that
\be \lim_{t\rightarrow \pm\infty} \alpha(t) = 0, \ee
making this trajectory asymptotically inertial, despite the drift.  As we shall now show, this solution has several analytically tractable results.  The special physical aspects of this solution will be investigated in the following sections. We shall refer to this horizonless mirror as Modified-BHC (``MBHC'') for short.  

\chapter{The Energy Production of MBHC} 

\label{sec:energy}

\section{The Energy Flux}

 The energy flux of a moving mirror was first derived by Davies and Fulling \cite{Davies:1976hi}.  Expressed in terms of the shock functions, it may be computed via
\bea
F(u) &=& \f{1}{24\pi}\left[\f{3}{2}\left(\f{v_s''}{v_s'}\right)^2-\f{v_s'''}{v_s'}\right],\label{stress}\\
F(v) &=& \f{-1}{24\pi} \left[ \f{3}{2}\left(\f{u_s''}{u_s'}\right)^2 - \f{u_s'''}{u_s'}\right]\f{1}{u_s'^2},\\
F(t) &=& \frac{1}{12\pi}\left[\f{x_s'''(x_s'^2-1)-3x_s'x_s''^2}{(x_s'-1)^4(x_s'+1)^2}\right],\label{stressz}\\
F(x) &=& \frac{1}{12\pi}\left[\f{t_s'''(t_s'^2-1)-3t_s't_s''^2}{(t_s'-1)^4(t_s'+1)^2}\right].
\eea

In terms of a $u$-dependent rapidity \cite{Good:2016oey}, $\eta(u) \equiv \tanh^{-1}[\dot{z(t_u)}] =\frac{1}{2}\ln v_s'(u)$, this is 
\be F(u) = \frac{1}{12\pi} \left(\eta'^2 - \eta''\right). \ee
\subsection*{Right Side}

\begin{figure}[ht]
\begin{center}
\mbox{\subfigure{\includegraphics[width=2.5in]{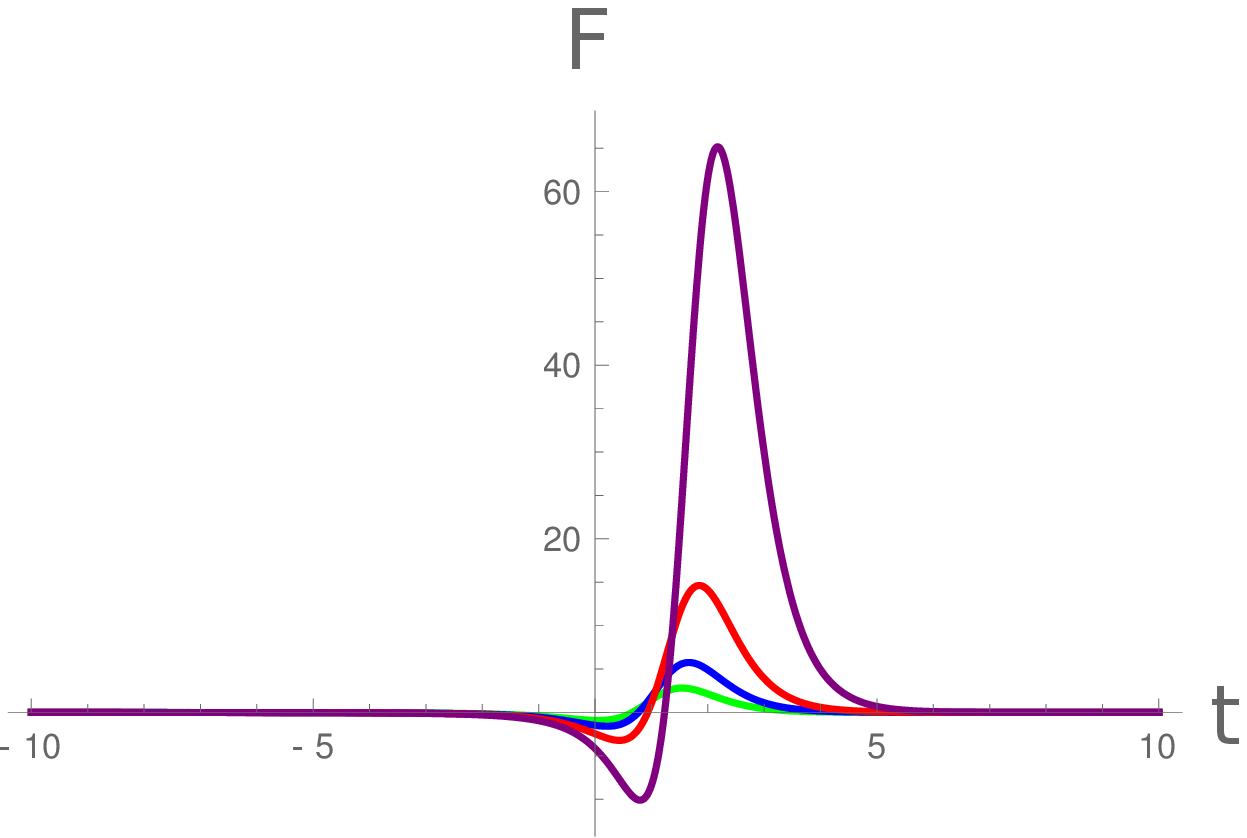}}\quad
\subfigure{\includegraphics[width=2.5in]{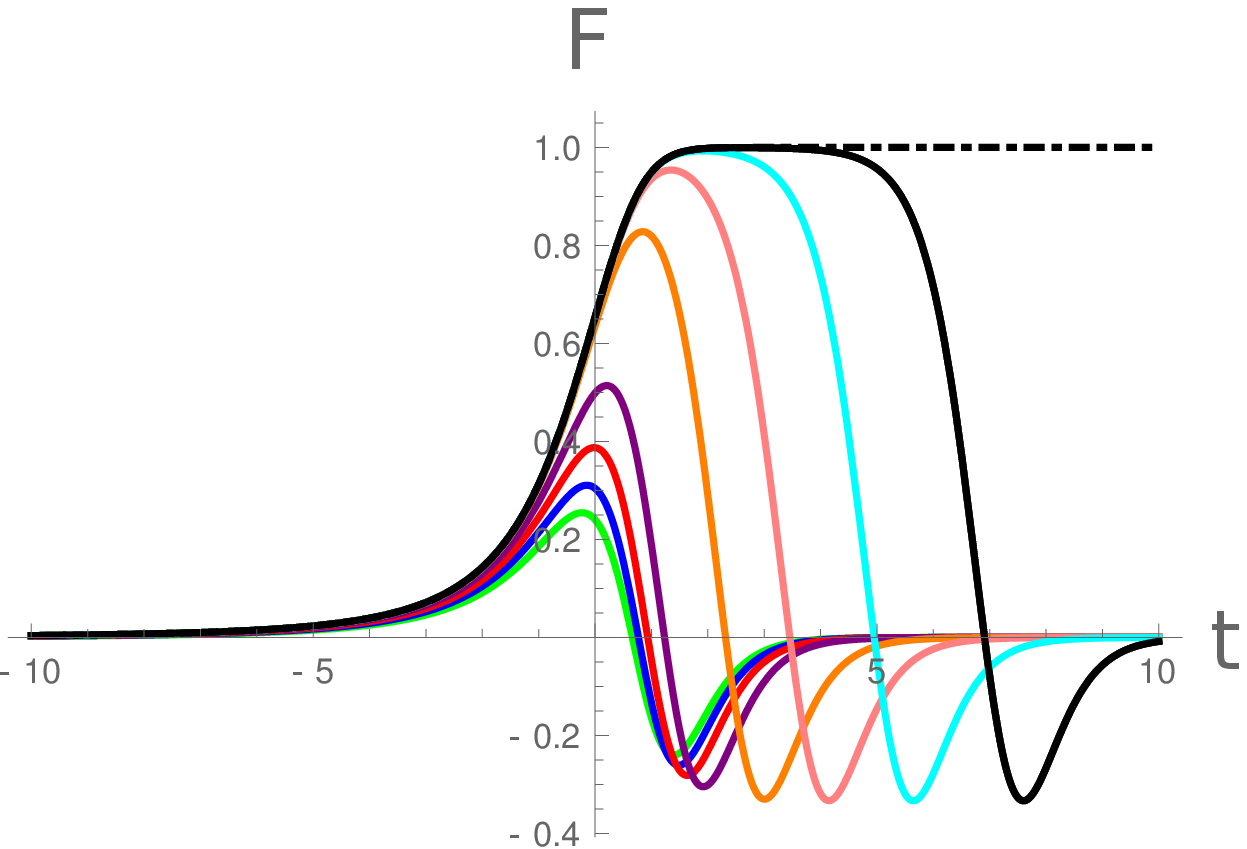} }}
\caption{\label{fig:flux} \textbf{Left:} The left observer sees energy flux that strongly peaks the faster the coasting speed of the mirror: $\xi = 0.6,0.7,0.8,0.9$, colored by green, blue, red, purple, respectively. Notice the initial NEF, and the above-thermal, $F(t)>1$, emission. Here $\kp^2 = 48\pi$.  \textbf{Right:} Successive plots of the energy flux observed by the right observer, $F(t)\equiv\la T_{uu} \ra$, Eq.~(\ref{energyflux}), from smallest peak to the largest peak, with varying limiting mirror speeds, $\xi=0.6$,$0.7$,$0.8$,$0.9$. Also included are $\xi=1-0.1^x$ where $x=2,3,4,6$ colored by orange, pink, cyan, black, respectively.  Thermal equilibrium occurs only for a very fast final coasting speed.  The total NEF is qualitatively unchanged at this speed.  The acceleration parameter is set to $\kp^2 = 48\pi$ so that the plateau levels out at $F=1$, the dot-dashed line. }
\end{center}
\end{figure}

The energy flux, emitted to an observer at the right side of the mirror, $\mathscr{I}_R^+$, using the trajectory of Eq.~(\ref{trajectory}) in Eq.~(\ref{stressz}), is therefore easily calculated:
\be \label{energyflux}F(t) =  \frac{\kappa ^2 \xi  W\left(2 e^{-2 \kappa  t}\right) \left(\xi ^2+2 W\left(2 e^{-2 \kappa  t}\right)^2+W\left(2 e^{-2 \kappa  t}\right)-1\right)}{3 \pi  \left(-\xi +W\left(2 e^{-2 \kappa  t}\right)+1\right)^2 \left(\xi +W\left(2 e^{-2 \kappa  t}\right)+1\right)^4}.\ee
It contains a build-up phase, a thermal plateau, and an end-phase accompanied by negative energy flux (NEF), see Figure (\ref{fig:flux}).  The residual horizon location has been set to $v_H=0$.

A period of thermal emission occurs at extremely high coasting speeds, giving a thermal plateau, which is, in the limit $\xi \rightarrow 1$, located for some time $\Delta t_{TP}$, at  
\be \label{plateau} F(\Delta t_{TP}) \approx F_T \equiv \f{\kp^2}{48 \pi}. \ee 
Interestingly, this is the same as the constant flux produced by the (eternally thermal) Carlitz-Willey trajectory \cite{Carlitz:1986nh}.  The Carlitz-Willey mirror radiates a thermal Planckian distribution of particles for all times, at $F = F_T$.  In our model, this value occurs because in the limit $\xi \rightarrow 1$, this mirror has the same shock functions as the black mirror, which has thermal radiation at late times.  However, now it is apparent that in this model, the evaporation eventually stops, effectively decoupling the late-time approximation from the high-frequency approximation.

Allowing $\xi$ to be nearly the speed of light, (for example, $\xi = 1-0.1^{10}$, with no formal limit), the energy flux emitted to the right observer, has a simple minimal negative value, at some late time, $t_0$, where 
\be F(t_0)^{\text{min}} = -\f{1}{3}F_T, \ee  
which is a fairly significant proportion of the maximum magnitude amplitude of thermal emission.        

\subsection*{Left Side}

The energy flux, emitted to an observer at the left of the mirror, $\mathscr{I}_L^+$, using the trajectory of Eq.~(\ref{trajectory}) by symmetry reversing the sign on $\xi$, is:
\be \label{energyfluxleft}F(t) = -\frac{\kappa ^2 \xi  W\left(2 e^{-2 \kappa  t}\right) \left(\xi ^2+2 W\left(2 e^{-2 \kappa  t}\right)^2+W\left(2 e^{-2 \kappa  t}\right)-1\right)}{3 \pi  \left(-\xi +W\left(2 e^{-2 \kappa  t}\right)+1\right)^4 \left(\xi +W\left(2 e^{-2 \kappa  t}\right)+1\right)^2}.\ee
The energy flux contains an initial nascent NEF, a rapid reversal to positive energy flux and build-up to a non-thermal positive energy flux peak, and finally a rapid end-phase that falls to zero emission, see Figure (\ref{fig:flux}).
It is now clear that while one side of the mirror is approaching thermal equilibrium emission, the other side is experiencing a single non-thermal, ever-more-narrow burst, demonstrating a characteristic difference between the left and right observers. We investigate the pulse via particle spectra in Section (\ref{sec:particles}).  

\section{Temperature of MBHC}
MBHC achieves a temperature, $2\pi T = \kappa$, to lowest order in $\epsilon$ where $\xi \equiv 1-\epsilon$, via a ``twice rapid acceleration'' ($\kappa(u) = |p''/p'| = |2\eta'|$) approximation.  The rapid acceleration, $\eta'(u)$, is identically constant, such that $\kappa(u) = \kappa$, for the eternally thermal mirror (Carlitz-Willey).  One finds, 
\be 2\pi T = |2\eta'|= \kappa(1+W(e^{-\kappa u}))^{-2} + \mathcal{O}(\epsilon). \ee
For large $\kappa u$, so long as, $\kappa u \lll \epsilon^{-1}$, then $W(e^{-\kappa u}) \rightarrow 0$, and to lowest order in $\epsilon$, the rapid acceleration is constant, $2\pi T = |2 \eta'| = \kappa$.
   
From the energy flux production, we can help quantify the equilibrium condition of MBHC.  The simplicity of the time-space function, $t_s(x)$, allows for analytic tractability.  Finding where the radiation is most near equilibrium amplitude, $F \approx F_T \equiv \kappa^2/(48\pi)$ is possible.  Using $v_H=0$, and $t_s(x) = -\kp^{-1} e^{2\kp x/\xi} - x/\xi$, one has the flux as a function of space:
\be F(x,\xi,\kp) = \frac{2 \kp^2 \xi  e^{\frac{2\kp x}{ \xi }} \left(8 e^{\frac{4\kp x}{ \xi }}+2 e^{\frac{2\kp x}{\xi }}+\xi ^2-1\right)}{3 \pi \left(-2 e^{\frac{2\kp x}{ \xi }}+\xi -1\right)^2 \left(2 e^{\frac{2\kp x}{ \xi }}+\xi +1\right)^4}. \ee
Maximizing $F(x,\xi,\kp)$ with respect to $x$, gives the spatial location, $x_0$, where the flux is maximum, $F(x_0, \xi, \kp) = F_{\rm{max}}(\xi,\kp)$. Since drift speed is high, then to lowest order in $\epsilon$, ignoring the imaginary component of this spatial locus, the real location is  
\be x_0 = \f{1}{6\kp}  \ln \f{\epsilon}{6} + \mathcal{O}(\epsilon^{1/3}). \ee
The maximum flux, to lowest order in $\epsilon$, is then
\be F(x_0,\xi,\kappa) = F_{\rm{max}} (\xi,\kappa) = \f{\kappa^2}{48\pi}\left[1-3 \sqrt[3]{6} \epsilon^{2/3}+\frac{25}{3}\epsilon + \mathcal{O}(\epsilon^{4/3})\right]. \ee
Following Davies \cite{Davies:1977yv} Eq. 3.10, or Walker \cite{Walker:1984vj} Eq. 5.10, 
we consider the property that the energy flux of a thermal trajectory has
\be F = \int_0^\infty \f{d\omega}{2\pi} \f{\omega}{e^{\omega/T}-1} = \f{\pi}{12} T^2. \ee
Temperature can be expressed as,
\be T(\xi,\kappa) = \sqrt{\f{12}{\pi} F_{\rm{max}}(\xi,\kappa)}, \ee
where we have taken the positive root, $T>0$.  To low order in $\epsilon$, the result is
\be T(\xi,\kappa) = \f{\kappa}{2\pi}\left[1-3\left(\f{3}{4}\right)^{1/3} \epsilon^{2/3} + \f{25}{6} \epsilon + \mathcal{O}(\epsilon^{4/3})\right]. \ee
The lowest order dependence on drift speed scales as $\sim(1-\xi)^{2/3}$, indicating, e.g. that speeds of $\xi = 1-0.1^9$ give a millionth part deviation from equilibrium temperature.  To ensure MBHC is very near equilibrium for an extended period of time, we will use speeds far faster while investigating the time dependence of particle production in Section (\ref{sec:particles}).

\section{Total Energy Produced by MBHC}

It proves possible to calculate the finite total emitted energy analytically. In terms of the shock wave functions, the total energy to the right side of the mirror, is computed via,
\bea
E &=& \int_{-\infty}^{\infty} F(u) d u,\\
E &=& \int_{-\infty}^{\infty} F(v) u_s' d v,\\
E &=& \int_{-\infty}^{\infty} F(t) (1-x_s')d t, \label{fluxtime}\\
E &=& \int_{+\infty}^{-\infty} F(x) (t_s' - 1)d x. 
\eea
or after integration by parts, where the boundary term is ignored due to asymptotic inertial character,
\bea
E &=& \f{1}{48\pi}\int_{-\infty}^{\infty}\left(\f{v_s''}{v_s'}\right)^2 d u,\\
E &=& \f{1}{48\pi}\int_{-\infty}^{\infty}\f{u_s''^2}{u_s'^3}  d v,\\
E &=& \f{1}{12\pi}\int_{-\infty}^{\infty}\f{x_s''^2}{(1+x_s')^2(1-x_s')^3} \;d t,\\
E &=& \f{1}{12\pi}\int_{-\infty}^{\infty}\f{t_s''^2}{(1+t_s')^2(1-t_s')^3} \;d x.
\eea
where $v_s\equiv v_s(u)$, $u_s\equiv u_s(v)$, $x_s \equiv x_s(t)$ and $t_s \equiv t_s(x)$.  The primes always mean derivatives with respect to the respective function variable.  For the mirror trajectory here with finite energy production we shall use the $x_s(t)$ integral over $d t$ and confirm it with quanta summing of particles in Section (\ref{sec:particles}), where the total emitted energy is $E = \int_0^{\infty}d\w \; \w \int_0^{\infty}d\w' \;|\beta_{\w\w'}|^2 $.

Note that by ``total'', we mean that it is the total amount of energy that \emph{only} the observer on the right side detects.  The mirror emits energy on both sides to two separate observers: left and right.  To find the energy emitted to the left, by symmetry, one can simply reverse the motion and compute the energy on the right side again.    

\subsection{Right Side}
The total energy radiated to $\mathscr{I}_R^+$ is therefore:
\be \label{rightsideenergy} E_R = \frac{\kappa  (3- \xi) \tanh ^{-1}(\xi )}{48 \pi  \xi ^2}-\frac{\kappa  (3+ 2\xi)}{48 \pi  (\xi^2 + \xi)}.\ee
MBHC does not result in the emission of infinite energy to the usual observer at $\mathscr{I}_R^+$.  Note that the solution here is monotonic for increasing coasting speed and never negative for $0<\xi<1$.  Here, the $\lim_{\xi\rightarrow 0} E_R = 0$, and the $\lim_{\xi\rightarrow 1} E_R = +\infty$. 

\subsection{Left Side}
For an observer to the left at $\mathscr{I}_L^+$ the total energy emitted is found by simply substituting, $\xi\rightarrow -\xi$, into $E_R$, 
\be E_L = \frac{\kappa  (3+ \xi) \tanh ^{-1}(-\xi )}{48 \pi  \xi ^2}-\frac{\kappa  (3- 2\xi)}{48 \pi  (\xi^2 - \xi)}. \ee
Again, the energy is finite as long as the speed is less than the speed of light.  The expression, $E_L(\xi)$ is a monotonic function of $\xi$.

\subsection{Both Sides} 
\begin{figure}[ht]
\begin{center}
\mbox{\subfigure{\includegraphics[width=2.5in]{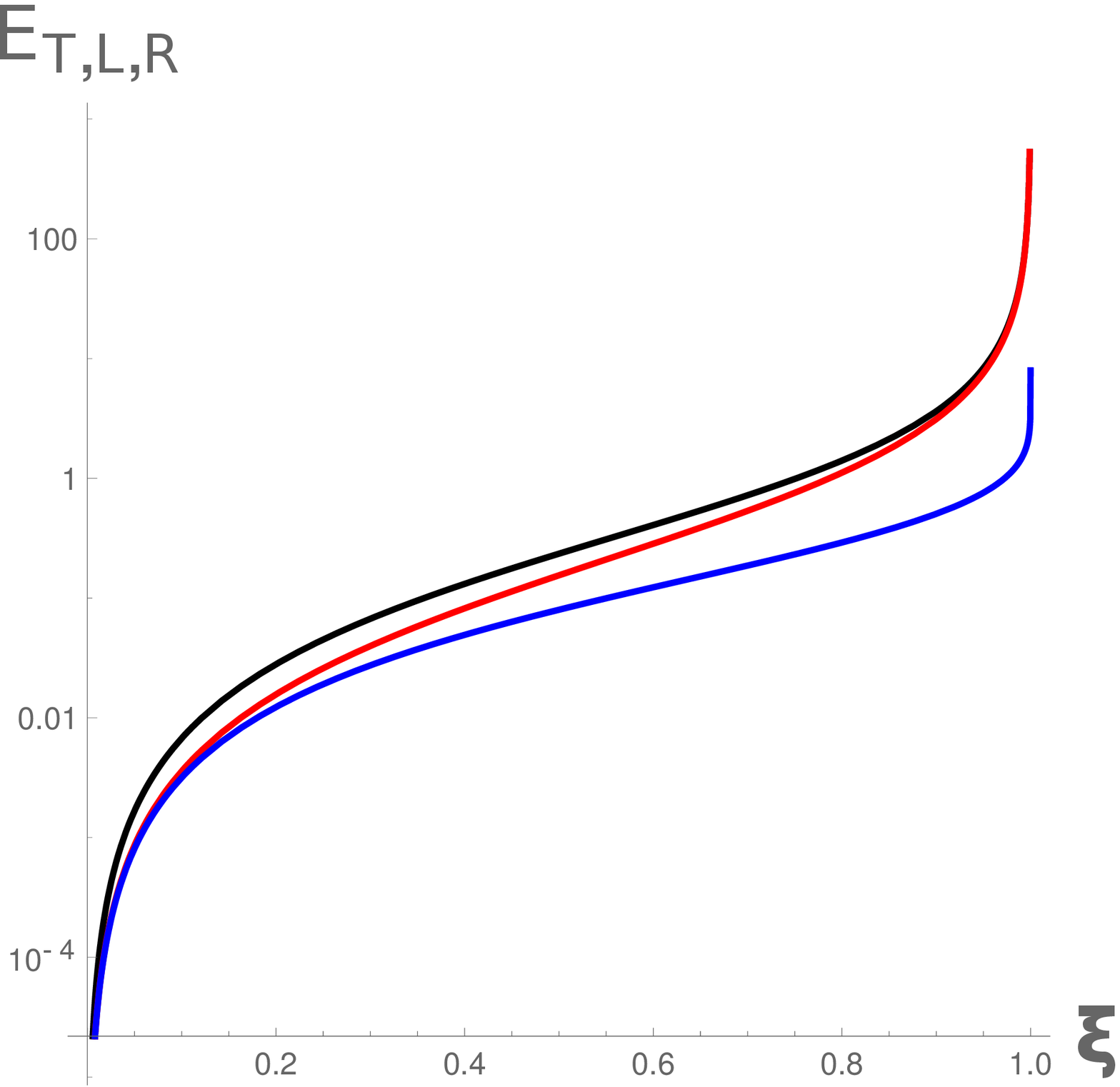}}\quad
\subfigure{\includegraphics[width=2.5in]{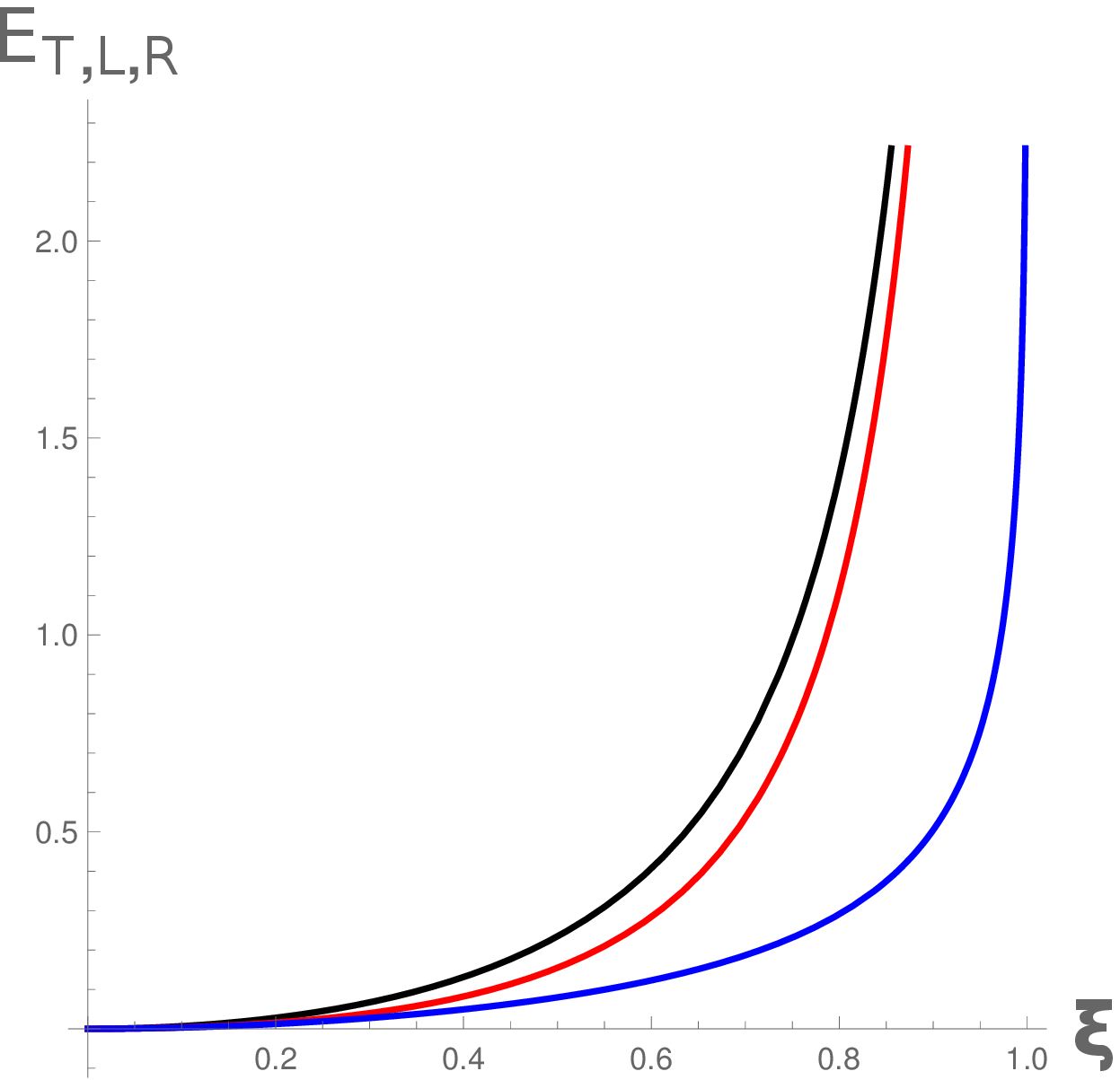} }}
\caption{\label{fig:totalenergy} \textbf{Left:}  Energies plotted in a semi-log plot.  The total energy emitted is the black line, $E_T(\xi) = \frac{\kp}{24\pi}(\frac{1}{1-\xi^2}-\frac{\tanh^{-1}\xi}{\xi})$, with $\kp = 24\pi$ in a log plot. The energy diverges as the final coasting speed approaches the speed of light. The red line is the $E_L$ and the blue line is $E_R$. \textbf{Right:} The same energies, $E_T$, $E_L$, and $E_R$, plotted to scale, $\kp = 24 \pi$. } 

\end{center}
\end{figure}

For the high coasting speeds we are interested in, the energy emitted to the left is \emph{always much} greater than the energy emitted to the right, $E_L\gg E_R$.  For small values of $\xi$ one finds
\be \frac{E_L}{E_R} = 1 + \frac{6}{5} \xi + \mathcal{O}(\xi^2), \ee
indicating that $E_L > E_R$.  As it turns out, $E_L > E_R$ for all values of the final drift speed, $0< \xi < 1$.    
The total energy emitted to both observers is $ E_T = E_L + E_R$:
\be E_T = \frac{\kappa}{24\pi}\left[\cosh ^2(\eta )-\eta  \coth (\eta )\right], \ee
where $\eta = \tanh^{-1} \xi$, is the final rapidity. See Figure (\ref{fig:totalenergy}) to see a graph of the combined total emitted energy from both sides of the mirror.  Notice the divergent behavior as the coasting speed approaches the speed of light.  The energy increases monotonically as a function of the coasting speed. 

\chapter{The Correlations in the Radiation} 

\label{sec:correlations} 

\section{Correlation Functions}

An under-appreciated lesson stressed by Ford and Roman \cite{Ford:2004ba}, is that there is a great deal more happening in the accelerating mirror geometry than is revealed by the expectation value of the stress-energy tensor alone. There are subtle increases or reductions in correlations between the flux along rays even where the expectation value vanishes.  The stress-energy tensor correlation function is of interest in our situation because it reveals information about the energy flux that demonstrates the thermal character of the radiation above and beyond that of the thermal plateau of the stress-energy tensor during equilibrium.  The shock functions for the moving mirror are needed to compute the correlation functions for the stress-energy tensor. 
It was previously shown that the ray-tracing function $p(u)$ is useful for delta-function pulse piece-wise mirror trajectories \cite{Ford:2004ba}.  In this section, we extend this work to continuous trajectories and compute the correlations with an emphasis on the equilibrium period of MBHC.  The energy fluxes emitted by any moving mirror can be positive and negative, but they are only average values.  The fluctuations around this average value are generally expected because the quantum state is not an eigenstate of the stress-energy tensor operator. 

The correlation function for the stress-energy tensor is
\be C_{\mu\nu \mu'\nu'} = \la T_{\mu\nu}(y)T_{\mu'\nu'}(y')\ra - \la T_{\mu\nu}(y)\ra \la T_{\mu'\nu'}(y')\ra ,\ee
where the spacetime points are indicated by $y=(u,v)$ and $y' = (u',v')$.  The correlation functions between two right moving rays, two left moving rays, and right and left moving rays are, respectively:
\be C_{RR}(u,u') = \la T_{uu}(u)T_{uu}(u')\ra - \la T_{uu}(u)\ra \la T_{uu}(u')\ra, \ee
\be C_{LL}(v,v') = \la T_{vv}(v)T_{vv}(v')\ra - \la T_{vv}(v)\ra \la T_{vv}(v')\ra,\ee
\be C_{LR}(v,u') = \la T_{vv}(v)T_{uu}(u')\ra - \la T_{vv}(v)\ra \la T_{uu}(u')\ra. \ee
Solved in terms of the ray tracing function, $p(u)$, the results are \cite{Ford:2004ba}
\be C_{RR}(u,u') = \f{[p'(u')p'(u)]^2}{8\pi^2[p(u')-p(u)]^4}, \ee
\be C_{LL}(v,v') = \f{1}{8\pi^2[v'-v]^4}, \ee
\be C_{LR}(v,u') = \f{[p'(u')]^2}{8\pi^2[p(u')-v]^4}, \ee
where $p'(u) = d p(u)/d u$ and $p'(u')=d p(u')/d u'$.  

The above expressions deal only with correlations of distinct rays.  These expressions simplify, as would be expected, in vacuum or with a static mirror present.  For a static mirror we have the condition, $v = p(u) = u$, and 
\be C_{RR}(u,u') = C_{\text{vac} \oplus \text{static}}(u,u') = \f{1}{8\pi^2[u'-u]^4}, \ee
\be C_{LL}(v,v') = C_{\text{vac} \oplus \text{static}}(v,v') = \f{1}{8\pi^2[v'-v]^4}, \ee
\be C_{LR}(v,u') = C_{\text{static}}(v,u') =  \f{1}{8\pi^2[u'-v]^4}. \ee
In vacuum $C_{LR}(v,u') = 0$ because there can only be correlations with left and right moving fluxes with a mirror present.  The correlation limits for $C_{RR}$ and $C_{LL}$ hold for either vacuum or a static mirror, hence the xor, $\oplus$, in the subscript.  The ratios 
\be \label{R1} R_1 \equiv \f{C_{RR}(u,u')}{ C_{\text{vac} \oplus \text{static}}(u,u') },\ee
and
\be \label{R2} R_2 \equiv \f{C_{LR}(v,u')}{ C_{\text{static}}(v,u')} \ee
can tell us about enhancement and suppression of correlations.  For $R_i>1$ one interprets enhancement, for $R_i<1$ there is suppression. 
\section{Correlation Solutions}

We focus on one ratio only in order to help confirm thermal equilibrium to the right observer.  This correlation ratio, $R_1$, involving $C_{RR}$, associates two right moving rays.  These rays come off the mirror heading to the right observer at $\mathscr{I}_R^+$.  
The $R_1$ solutions for the ratios for the three mirrors of interest,
\begin{enumerate}
	\item[(1)] Thermal Mirror (Carlitz-Willey)
	\item[(2)] Black Mirror (BHC)
	\item[(3)] Horizonless Mirror (MBHC)
\end{enumerate}
are included here for completeness.  They are respectively,
\be R_1^{\textrm{Thermal}} = \frac{\kappa ^4 (u-u')^4 e^{2 \kappa  (u+u')}}{\left(e^{\kappa  u}-e^{\kappa  u'}\right)^4}, \ee
which, illustrates thermal correlations at all times.  However, for the black mirror, we have
\be R_1^{\textrm{BHC}} = \frac{\kappa ^4 (u-u')^4 W\left(e^{-\kappa u}\right)^2 W\left(e^{-\kappa u'}\right)^2}{\left(W\left(e^{-\kappa u}\right)+1\right)^2 \left(W\left(e^{-\kappa u'}\right)+1\right)^2 \left(W\left(e^{-\kappa u}\right)-W\left(e^{-\kappa u'}\right)\right)^4}, \ee
which, illustrates thermal correlations at late times.  Finally, we write down the MBHC's ratio, 
\be R_1^{\textrm{MBHC}} = \frac{\kappa ^4 (u-u')^4 \left(-\xi +(\xi +1) W_u+1\right)^2 \left(-\xi +(\xi +1) W_{u'}+1\right)^2}{\left(W_u+1\right)^2 \left(W_{u'}+1\right)^2 \left(\kappa  (\xi -1) (u-u')+\xi  (\xi +1) W_u-\xi  (\xi +1) W_{u'}\right)^4},\ee
where $W_u \equiv W\left(\frac{2 e^{-\frac{2 \kappa  u}{\xi +1}}}{\xi +1}\right)$, and  $W_{u'} \equiv W\left(\frac{2 e^{-\frac{2 \kappa  u'}{\xi +1}}}{\xi +1}\right)$.  

\begin{figure}[ht]
\begin{center}
\mbox{\subfigure{\includegraphics[width=2.5in]{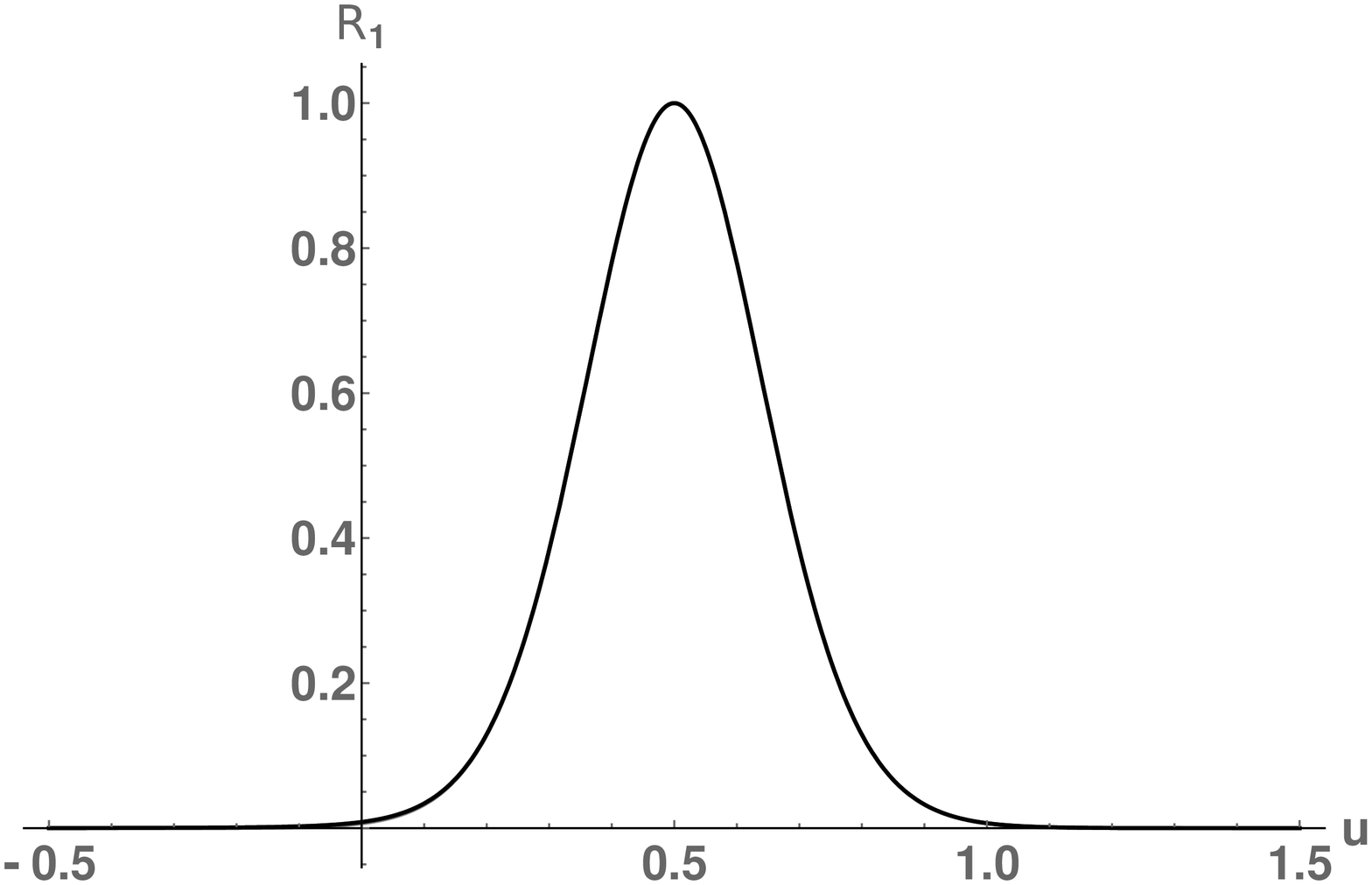}}\quad
\subfigure{\includegraphics[width=2.5in]{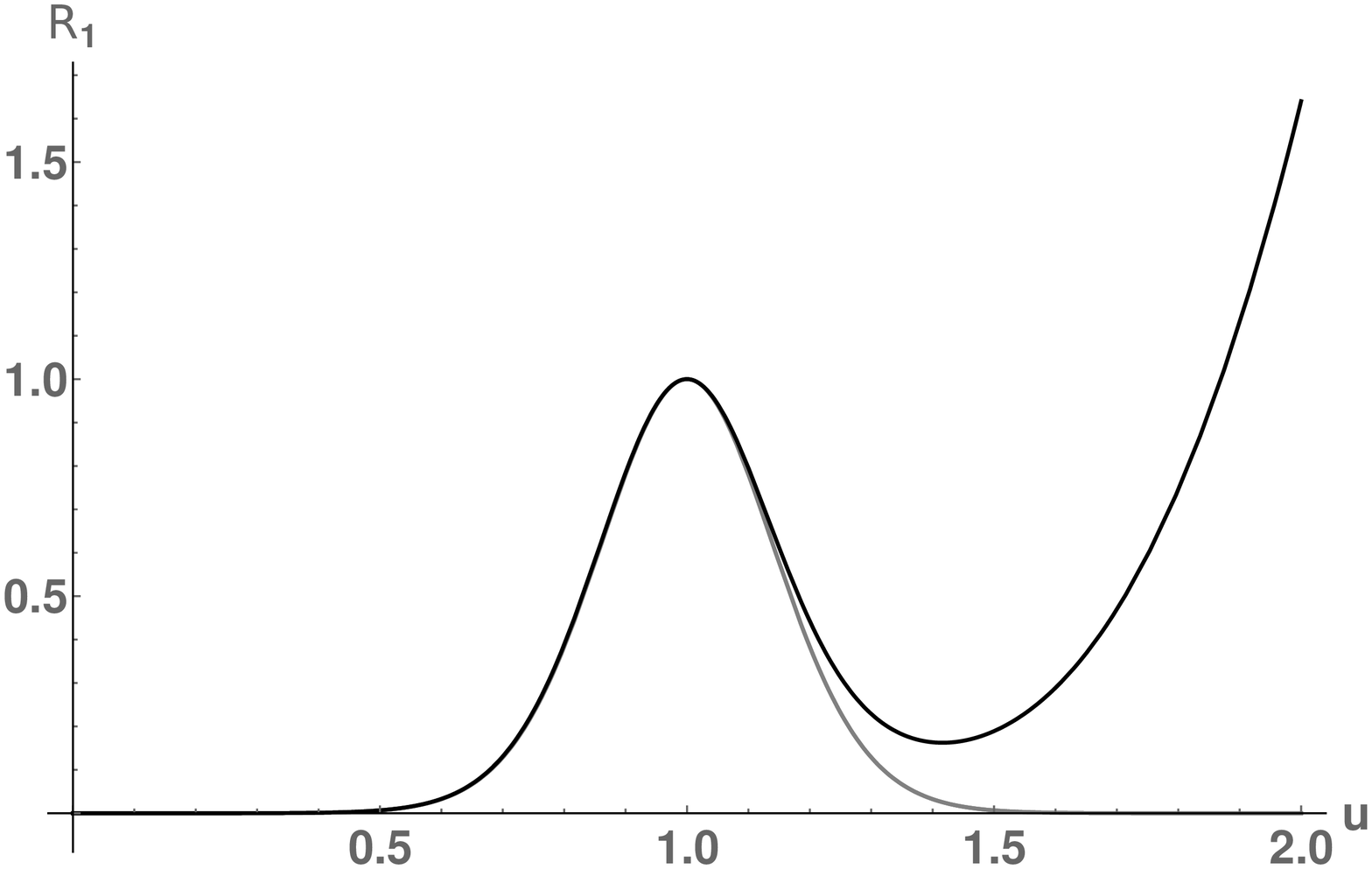} }}
\caption{\label{fig:R1thermal} \textbf{Left:} The ratio $R_1$ for all three mirrors: Carlitz-Willey, BHC and MBHC ($\xi = 1-10^{-7}$), where the right moving ray $u' = 0.5$ (chosen for simplicity) and $\kappa^2 = 48\pi$ (chosen to normalize a thermal energy flux of $F_T = \kappa^2/48\pi = 1$).  Notice all three correlations stack up closely on each other, indicating no difference in the correlations from thermal emission between all three mirrors.  With this $\kappa$ scale the maximum flux is near the ray $u'_0 = 0.48 \approx 0.50$ for MBHC.  \textbf{Right:} The ratio $R_1$ for all three mirrors where the right moving ray $u' = 1$ and $\kappa^2 = 48\pi$.  The gray line is the thermal mirror and BHC mirror.  The black line is MBHC.  Now we are comparing late time correlations, so late in fact, that MBHC has correlations that now deviate from thermal equilibrium.  This deviation signals the non-equilibrium completion of evaporation. Notice all three correlations stack up closely on each other only for early values of $u$, indicating no difference in the correlations from thermal emission for these rays.  The ratio becomes enhanced, $R_1>1$, once negative energy radiation is emitted.  } 
\end{center}
\end{figure}

All three mirrors give the same thermal correlations when comparing a ray that occurs in the appropriate equilibrium period. This is not at very late times for MBHC.  When one picks a very late time ray, then BHC and Carlitz-Willey are still thermal, but MBHC begins to break pattern to abide by the inevitable non-equilibrium completion of emission.  This is to be expected because at very late times the mirror abandons thermal character as the radiation ceases. See Figure (\ref{fig:R1thermal}).

\chapter{The Particle Production of MBHC} 

\label{sec:particles} 

\section{The Beta Bogoliubov Coefficient Integrals}

\begin{figure}[!h]
\centering
\mbox{\subfigure{\includegraphics[width=2.5in]{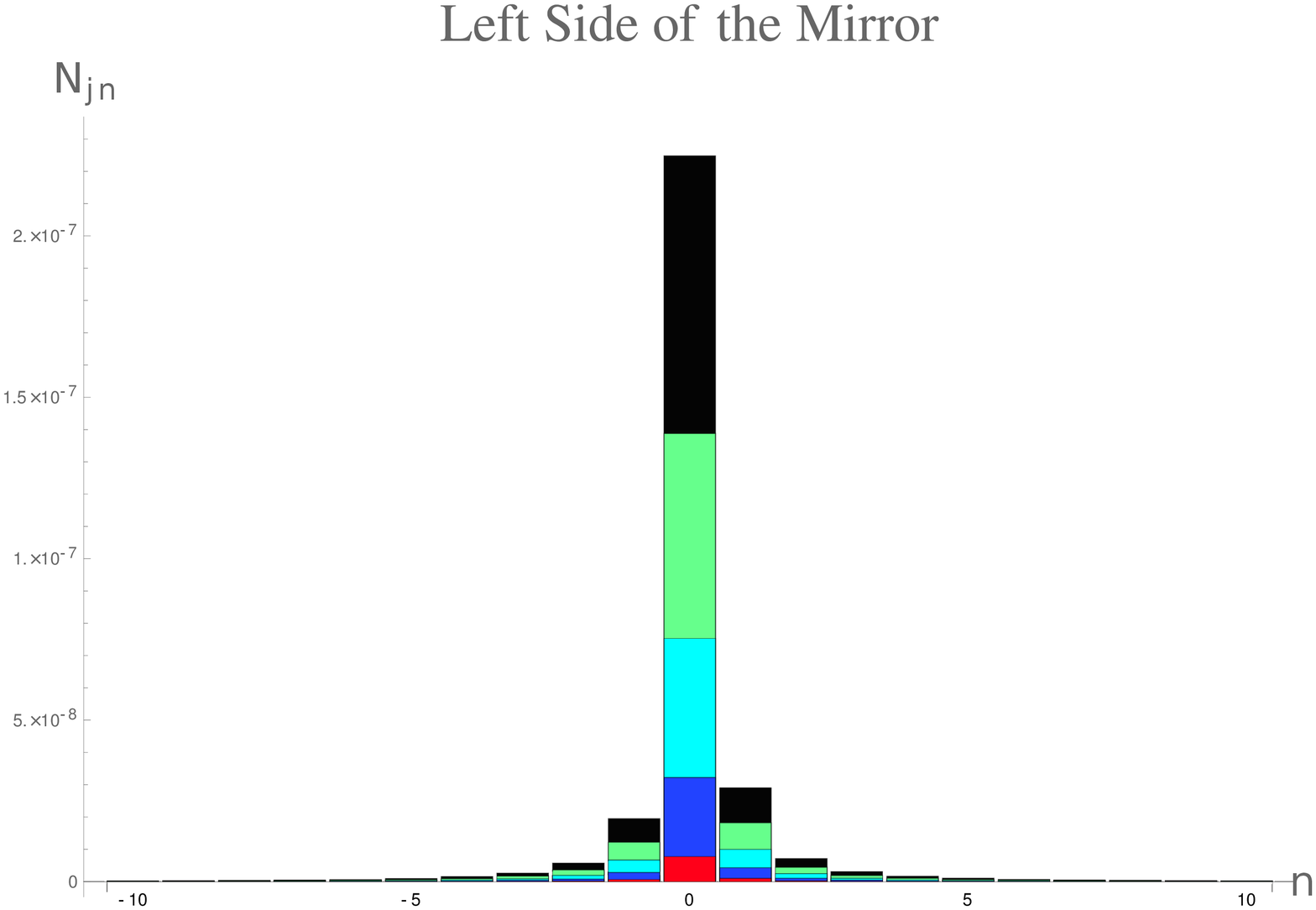}}\quad
\subfigure{\includegraphics[width=2.5in]{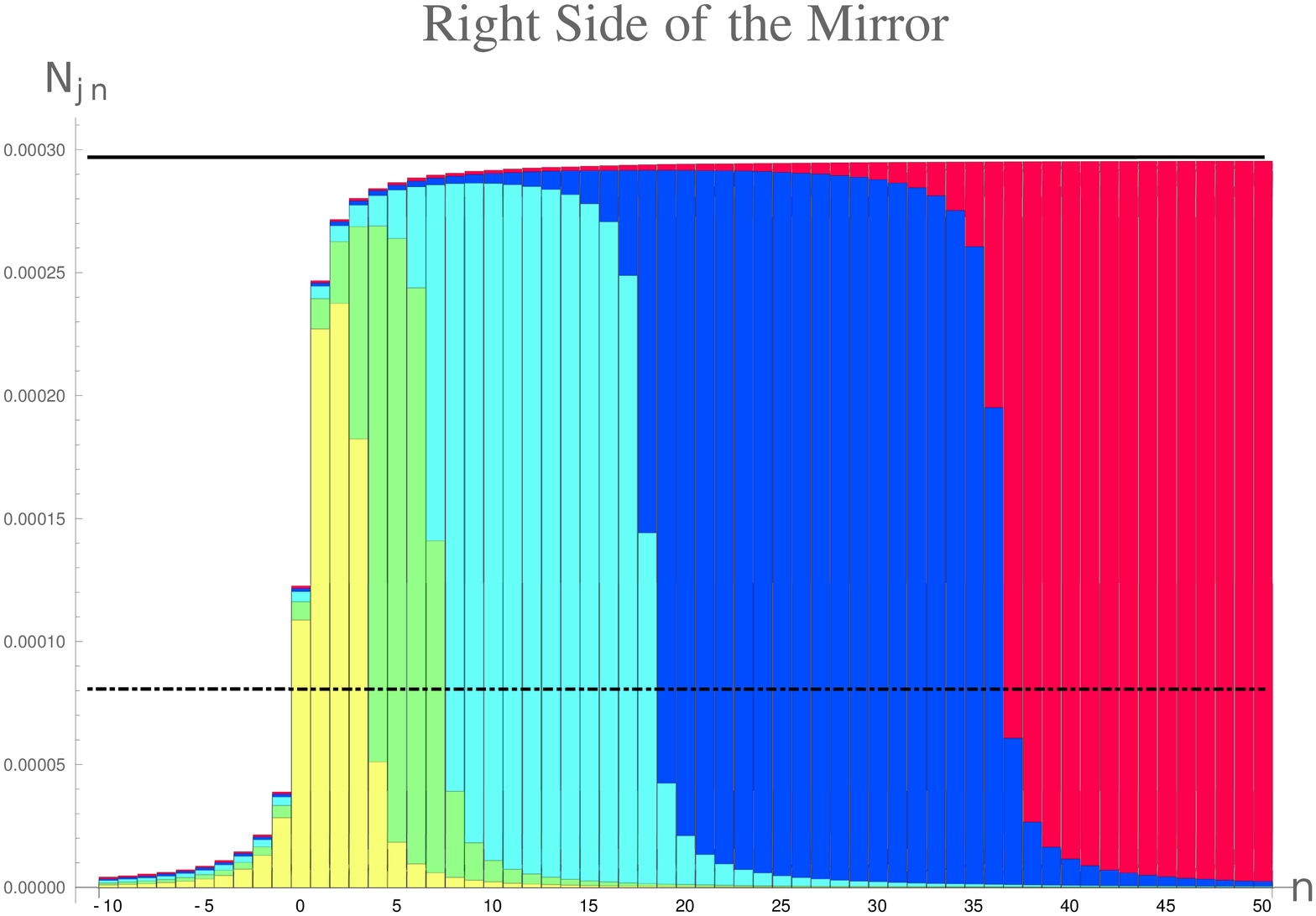} }}
\caption{\textbf{Left:} The left observer sees a non-thermal pulse of particle production.  The final drifting speed of MBHC need not be very high to see the growth of the pulse.  Here $\xi = 0.01 a $ where $a = 1,2,3,4,5$.  The acceleration parameter is set to $\kappa = 1$ and the lowest frequency bin $j=1$ is observed. Here $\epsilon = 1$.  The pulse of particles grossly exceeds constant Planckian emission with fast speeds. \textbf{Right:} The right observer sees a plateau of particle production.  The final drifting speed and other parameters are also $\xi = 1-0.1^{x}$, $\kappa=1$, $j=1$, and $\epsilon = 1$, where $x=10,20,50,100$.   Increasing the final coasting speed produces a flatter, more extended in time, thermal plateau. The particle production colored red is BHC, which is thermal at late times ad-infinitum.  The dotted-dashed line is the approximate Planck distribution, $N_j = (e^{\omega_j/T}-1)^{-1}$ where $\omega_j = (j+1/2)\epsilon$. The solid black line is the exact Planck distribution, $N_j = T\epsilon^{-1} \ln \left[ \frac{e^{(j+1)\epsilon/T}-1}{e^{j\epsilon/T}-1}\right] -1$, see e.g. \cite{Good:2016oey}, \cite{Good:2016bsq}. \label{fig:ppfast}} 
\end{figure}

The distinction between energy flux and particle flux has been studied by Walker-Davies \cite{Walker:1982} and Walker \cite{Walker:1984vj}. THese are great references.  The key ingredient is the information of the mirror trajectory equation of motion, which is encapsulated in the shock functions. One needs the Bogoliubov coefficients because the particle emission detected by an observer is
\be \la N_{\omega} \ra \equiv \la 0_{\rm{in}} | N^{\rm{out}}_\omega | 0_{\rm{in}} \ra = \int_0^{\infty} |\beta_{\omega \omega'}|^2 \; d \omega'. \label{particlecount} \ee
 There are four ways to calculate the beta Bogoliubov coefficient using the shock functions: 

\be
\label{betap}
\beta_{\w\w'} = \f{1}{4\pi\sqrt{\w\w'}}\int_{-\infty}^{\infty} d u \;
e^{-i\w  u - i\w'v_s(u)} \left(\w' \frac{d v_s(u)}{d u} - \w \right) \;,
\ee

\be
\label{betaf}
\beta_{\w\w'} = \f{-1}{4\pi\sqrt{\w\w'}}\int_{-\infty}^{\infty} d v \;
e^{-i\w' v -i\w u_s(v)} \left(\w \frac{d u_s(v)}{d v} -\w' \right) \;,
\ee

\be
\label{betaz}
\beta_{\w\w'} =
\f{1}{4\pi\sqrt{\w\w'}}\int_{-\infty}^{\infty} d t \;
e^{-i\w_{p}t + i\w_{n} x_s(t)} \left(\w_{p}\frac{ d x_s(t)}{d t}-\w_{n}\right) \;,
\ee
\be \beta_{\w\w'} =
\f{-1}{4\pi\sqrt{\w\w'}}\int_{+\infty}^{-\infty} d x \;
e^{i\w_{n} x -i\w_{p}t_s(x)} \left(\w_{n}\frac{d t_s(x)}{d x} - \w_{p} \right) \label{betat}\;.
\ee
Here $\w_p \equiv \w + \w'$ and $\w_{n} \equiv \w-\w'$. The integration bounds also assume all light rays hit the mirror and propagate to future null infinity on the right.  For light rays that do not hit the mirror, one must stop short the integration and add up only to the last null ray as described by the relevant variable.  All of the bounds are written for a mirror which starts at positive spatial infinity and proceeds left to negative spatial infinity in the far future.  The bounds must be appropriately changed for mirrors which have different behaviors and/or horizons.  For MBHC we use Eq.~(\ref{betat}) because of the simplicity of $t_s(x)$ function particular to MBHC. Notice the negative sign analogous to Eq.~(\ref{betaf}) drops away because the mirror starts at $x \rightarrow +\infty$ at $t \rightarrow -\infty$.  This integral can be used to obtain other trajectories when the other integrals are intractable.  We choose this integral because it will allow investigation of particle production while avoiding integration with the product log, (see \cite{Good:2013lca} for $z(t)$ approach), and therefore insert MBHC's trajectory, Eq.~(\ref{trajectory}), into the integral Eq.~(\ref{betat}) identifying, $t_s(x) = t(z)$. 
The solution is 
\be \label{particlebeta} \beta_{\w\w'\xi} = -\frac{\xi  \sqrt{\w \w'} (-i (\w+\w')/\kp)^{-A} }{2 \pi \kp (\w+\w')}\Gamma \left(A\right),\ee
with $A\equiv \frac{i}{2\kp} ((1+\xi)\w+(1-\xi)\w')$. 
This exact beta solution also contains the thermal plateau, similar to the instantaneous energy flux which closely approaches the thermal line at $F = F_T$.  Both particle and energy flux approach the thermal plateau only at high final speeds making it a salient feature of the radiation.  One should be confident the behavior shows up in both the particle production and energy flux because the packetized particles carry signatures of the instantaneous energy flux emission \cite{Good:2015nja}.  
In Figure (\ref{fig:ppfast}), we construct the usual localized beta Bogoliubov coefficients from the global beta Bogoliubov coefficients in Eq. (\ref{particlebeta}),
\be
\beta_{jn\w'} =
\f{1}{\sqrt{\epsilon}}\int_{j\epsilon}^{(j+1)\epsilon}
d\w \; \left[e^{\frac{2\pi i \w n}{\epsilon}} \beta_{\w\w'\xi} \right]\;.
\label{beta-packet}
\ee
These are the usual orthonormal complete wave packets \cite{Hawking2} which are used to find the time-frequency localized particle count,
\bea  
\la N_{jn}\ra &=& \int_0^\infty d\w' |\beta_{jn,\w'}|^2   \;,  
\nonumber 
\\
&=& \int_0^\infty d\w' \int_{j \epsilon}^{(j+1)\epsilon} 
\frac{d \w_1}{\sqrt{\epsilon}} \int_{j \epsilon}^{(j+1)\epsilon} 
\frac{d \w_2}{\sqrt{\epsilon}} \left[ e^{\frac{2 \pi i(\w_1- \w_2)n}{\epsilon}}
\beta_{\w_1 \w'} \beta^{*}_{\w_2 \, \w'} \right] \;. 
\label{Njn} 
\eea
 Particles arrive at $\mathscr{I}^+_R$ in the range of frequencies $j \epsilon \leqslant \omega \leqslant (j+1) \epsilon$ and in the range of times $ (2\pi n - \pi)/\epsilon \lesssim u \lesssim (2 \pi n + \pi)/\epsilon$.  Details on how to construct these packets in more general situations can be found in \cite{Good:2013lca}.  

\section{Global Particle Distribution}
The horizonless beta solution Eq.~(\ref{particlebeta}) of MBHC has distribution,
\be |\beta_{\w \w' \xi}|^2 = \frac{\omega'}{\pi  \kappa  \left(\omega' + \omega \right)^2}\frac{\xi ^2 \omega }{\left((1+\xi) \omega + (1 -\xi ) \omega '\right)}\frac{1}{ \left(e^{\frac{\pi}{\kappa}  \left((1+\xi) \omega + (1-\xi) \omega'\right)}-1\right)},\ee
while the horizon beta Bogoliubov coefficients of BHC have particle count per mode-squared, 
\be |\beta_{\w \w'}|^2 = \frac{\w'}{2\pi \kappa (\w+\w')^2}\frac{1}{e^{2 \pi \w /\kappa}-1}. \label{BHCdistribution} \ee
It is easy to see that as $\xi \rightarrow 1$ that the spectra coincide. The high frequency approximation, $\w'\gg \w$, applied to BHC's distribution, Eq.~(\ref{BHCdistribution}), gives the usual thermal result of Carlitz-Willey and late-time Davies-Fulling trajectories,
\be|\beta_{\w \w'}|^2_{T} = \frac{1}{2 \pi  \kappa  \w'}\frac{1}{ e^{2 \pi  \w/\kappa }-1}.\ee
One checks that the MBHC distribution is thermal, $|\beta_{\w \w'\xi}|^2 \approx |\beta_{\w \w'}|^2_T$, by a series expansion that first approximates the distribution with a very fast end-state drifting mirror $\xi \approx 1$ and then applies the high-frequency approximation $\w'\gg \w$.  This approach explicitly decouples the late-time approximation from the high-frequency approximation.

\section{Consistency Check}
It is a fair claim that the total summation of the energies of each quanta should be equal to the integral over the energy flux:
\be \int_0^\infty \omega \la N_\omega \ra d\omega = \int_{-\infty}^{\infty} F(u) d u. \ee 
Therefore, the beta Bogoliubov particle results can be confirmed to be consistent with the stress-energy by computing the total energy, using Eq.~(\ref{particlecount}) and Eq.~(\ref{fluxtime}),
\be \label{quantasum} \int_0^\infty \omega \left[\int_0^\infty |\beta_{\omega\omega'}|^2 d\omega'\right] d\omega = \int_{-\infty}^{\infty} F(t) (1-\dot{z})d t. \ee 
This consistency helps confirm the particles do indeed carry the energy.  
Explicitly, we have numerically confirmed the beta coefficients, Eq.~(\ref{particlebeta}), that for an observer to the right at $\mathscr{I}_R^+$, the total energy emitted is 
\be E_R = \int_0^\infty \int_0^\infty \w \cdot |\beta_{\w \w' \xi }|^2 \; d\w \; d\w' = \frac{\kappa  (3- \xi) \tanh ^{-1}(\xi )}{48 \pi  \xi ^2}-\frac{\kappa  (3+ 2\xi)}{48 \pi  (\xi^2 + \xi)},\ee
and for an observer to the left at $\mathscr{I}_L^+$, the total energy emitted is 
\be E_L = \int_0^\infty \int_0^\infty \w' \cdot |\beta_{\w \w' \xi }|^2 \; d\w \; d\w' = \frac{\kappa(3+ \xi) \tanh ^{-1}(-\xi )}{48 \pi  \xi ^2} -\frac{\kappa  (3-2\xi)}{48 \pi  (\xi^2 - \xi)},\ee
in agreement with the analytical results of the stress-energy tensor of Section (\ref{sec:energy}).  We have found the largest relative numerical error here to be less than $10^{-11}$ using $\kappa =1$, and various values of $0< \xi <1$.   


\chapter{Discussions}

\label{sec:conclusions} 

In the Introduction we raised the following question: ``Is there a moving mirror in (1+1)-dimensions, satisfying unitarity in the sense allowed by the Bianchi-Smerlak criterion (namely, $S(u) \to \text{const.}$ as $u \to \pm\infty$), that has no acceleration horizon, produces finite amount of energy, and serves as a limiting case analog for Eddington-Finkelstein coordinate null shell gravitational collapse?'' We have answered this question in the affirmative, by constructing an exact mirror solution that satisfies these properties. 
Furthermore, we investigated both sides of the mirror trajectory, and found interesting features regarding negative entropy and negative energy flux.

The hallmark trait of the solution is the fact that it is an asymptotically coasting mirror which does not have an accelerating horizon, yet approaches arbitrarily close to thermal equilibrium.  Thermal radiation arises from a sufficiently fast final drifting speed. The ray-tracing function is identical to the spacetime matching condition of the black hole case in the limit that the mirror drifts to the speed of light.

The global approach to treating horizons tends to work well in fully equilibrium thermodynamics, especially so with \emph{a priori} non-dynamical assumptions (i.e. constant energy flux) \cite{Carlitz:1986nh}.  It is well-known that the non-equilibrium cases are not so easy to formulate using the traditional methods.  \emph{A practical outcome of this thesis has been to show how robust the traditional methods can be when the horizon is removed from the start.  Non-equilibrium dynamical conditions follow suit, however the system can still achieve equilibrium, for an arbitrary extended amount of time.} With a consistent scaling ($\kappa$ in MBHC and BHC are the same scaled parameter relative to thermal emission), we have explicitly used the global geometric properties of the spacetime, and in the case of particle creation, only localized \textit{after} solving for the global beta Bogoliubov coefficients.   

The new mirror was described in terms of its energy flux, total energy, entropy flux, correlations, and particle flux.  The temperature can be detected by asymptotic observers with particle detectors (the radiation demonstrates a Planckian distribution for a very fast final drift speed and the use of the high frequency approximation).  The evidence for thermality is strengthened further by the long-lived steady-state stress-energy tensor and the correlations which match the eternally thermal equilibrium mirror of Carlitz-Willey \cite{Carlitz:1986nh}. 
However, a few remarks are in order comparing our analysis to Carlitz-Willey's seperate 1987 trajectory \cite{Carlitz1987} which is not eternally thermal.  A notable similarity between this apparent horizon trajectory and the one presented here is the constant rate of particle emission during a finite period of time. We confirm the locally thermal state in both unitary moving mirror trajectories.  One notable difference is that their trajectory is in terms of an approximate ray-tracing function with a kink.\footnote{Carlitz and Willey comment that the kink can be smoothed out but it is not done because it is clear it would not affect the conclusions much.}  The trajectory in this paper is exact, $C^\infty$, for all-times, and is expressed as an explicit space-time trajectory function, Eq.~(\ref{trajectory}).  A consequence of this fact is that it so happens our trajectory does not come to rest at late times, while their trajectory requires the mirror eventually become stationary and consequently the entire remnant mass is radiated away to leave behind a flat region of spacetime.  In other words, Carlitz-Willey consider a ``meta-stable" or ``long-lived" remnant that slowly evaporates away, whereas our remnant is eternal.

Unlike the eternal thermal Carltiz-Willey mirror \cite{Carlitz:1986nh} or the black mirror \cite{Good:2016oey}, the MBHC mirror gives rise to negative energy flux, and by the result of Bianchi-Smerlak \cite{Bianchi:2014vea,Bianchi:2014qua}, also to the non-monotonic mass loss of the any corresponding black hole.  Current efforts are being directed to explore the generalization of the tortoise coordinate from 
\be r^* \equiv r + 2 M \ln \left(\frac{r}{2 M}-1\right)\ee
to
\be r^*(\xi) \equiv r+ 2 M \xi  \ln \left[\frac{1-\xi}{2} \mathcal{W}\left(\frac{2 e^{\frac{r-2 M}{M (1-\xi)}}}{1-\xi}\right)\right], \label{tort} \ee
which can be evident from the generalization of the shock function of Eq.~(\ref{matchingfunction}) to Eq.~(\ref{matchingfunction2}), in the null-shell case which matches spacetimes outside and inside the shell.  For details on the null shell case, see e.g. Unruh (1976) \cite{Unruh:1976db}, Massar (1996) \cite{Massar:1996tx} or Fabbri (2005) \cite{Fabbri:2005mw}.  The generalization Eq.~(\ref{tort}) and a possible coupling between the parameters $\xi$ and $M$, may provide clues to understanding any corresponding black hole solution of MBHC, and by necessity a different all-time collapse scenario.  It is understood \cite{Hawking2} that at very early times of gravitational collapse, the system cannot be described by the no-hair theorem.  Therefore it is appropriate to consider the type of modifications that can provide various early time approaches to a thermal distribution, particularly those modifications that can afford unitarity and finite evaporation energy. 
The modifications that can take into account energy conservation like those of the dilaton gravity models have had significant success as a laboratory for studying black hole evaporation. The physical problem in 1+1 dilaton gravity of the evaporating black hole and its modified emission extends to complete evaporation for the Russo, Susskind, and Thorlacius (RST) model and to partial evaporation leaving a remnant for the Bose, Parker, and Peleg (BPP) model .  The similarity of the MBHC mirror to the BPP model is striking in several qualitative aspects: NEF emission as a thunderpop, a left over remnant, and finite total energy emission.  It is also interesting that the mass of the remnant in the BPP model is independent of the mass $M$ of the infalling matter, since with respect to the issue of energy conservation, there is no known physical analog for $M = 1/(4\kappa)$, the initial mass of the shockwave, in the mirror model.

Finally, we shall comment on the peculiar emission we find on the left side of the mirror trajectory, in particular, its possible relevance to the black hole correspondence (if one exists). We conjecture that the ``left emission'' corresponds to in-falling flux into the black hole. As such, its associated temperature and entropy might shed some light on the information paradox, since as we emphasized in the Introduction, unitarity is a property of the Hilbert space defined on the entire spacetime. In fact, such a ``left temperature'' in the context of black hole physics already exists in the literature, see e.g., \cite{1604.00465}. It should also be emphasized that the recent result in the literature \cite{1603.01964}, concerning the study of two-dimensional model of gravitational collapse, shows that a geodesic observer on the left side measures late time thermal radiation but zero flux. This result is drastically different from ours. It might be interesting to conduct a comparative study between our model with that of \cite{1603.01964}.

Ultimately, while MBHC is elementary, it embraces several surprisingly interesting traits.  Since some of these traits are shared with more sophisticated systems, this solution may be a precursor for ensuing developments (the overt example being curved spacetime collapse).  On the other hand, this solution has exposed several explicit general attributes which are unanticipated and must be understood in order to claim a good grasp of the dynamics of the particle creation effect in non-thermal equilibrium.  

The outstanding advantage of this mirror solution is the exact expressions for quantities of interest.  Since one natural speculation is the direct applicability to a curved spacetime analog, we aim to examine this pertinent and interesting follow-up topic in later manuscripts with primary consideration to energy conservation of the black hole's modified evaporation emission, metric continuity across the shock boundary, and to the Bogolubov coefficients of specific dilaton gravity models.  The soft-particle production problem may also be investigated along these lines. 




\printbibliography

\end{document}